\documentclass[preprint,longabstract]{aastex}

\usepackage[T1]{fontenc}
\usepackage[bitstream-charter]{mathdesign}
\usepackage{booktabs}
\usepackage[latin1]{inputenc}							
\usepackage{amsmath}									
\usepackage{hyperref}
\title{Understanding breaks in solar flares x-ray spectra: Evaluation of a co-spatial return-current model\
       \vspace{0.5cm}}

\author{%
	Meriem Alaoui, Gordon D. Holman \\
	Catholic University of America and NASA Goddard Space Flight Center \\
	meriem.alaouiabdallaoui@nasa.gov \\
	\vspace{20pt}
	}

\usepackage{graphicx}
\usepackage{natbib}
\begin{document}




\begin{abstract}
Hard x-ray spectral breaks are explained in terms of a 1D model with a co-spatial return current. We study 19 flares observed by RHESSI (Ramaty High Energy Solar Spectroscopic Imager) with strong spectral breaks at energies around a few deka-keV, that cannot be explained by isotropic albedo or non-uniform ionization alone. We identify these breaks at the HXR peak time, but we obtain 8 s-cadence spectra of the entire impulsive phase. Electrons with an initially power-law distribution and a sharp low-energy cutoff lose energy through return-current losses until they reach the thick target, where they lose their remaining energy through collisions. Our main results are: (1) The return-current collisional thick-target model (RCCTTM) provides acceptable fits for spectra with strong breaks. (2) Limits on the plasma resistivity are derived from the fitted potential drop and deduced electron-beam flux density, assuming the return-current is a drift current in the ambient plasma. These resistivities are typically 2-3 orders of magnitude higher than the Spitzer resistivity at the fitted temperature, and provide a test for the adequacy of classical resistivity and the stability of the return current. (3) Using the upper limit of the low-energy cutoff, the return current is always stable to the generation of ion acoustic and electrostatic ion cyclotron instabilities when the electron temperature is lower than 9 times the ion temperature. (4) In most cases the return current is most likely primarily carried by runaway electrons from the tail of the thermal distribution rather than the bulk drifting thermal electrons. For these cases, anomalous resistivity is not required.
\end{abstract}

%
\keywords{Flares $\cdot$ Flare X-Ray Observations $\cdot$ Return Currents in corona $\cdot$ Current-driven instabilities}


\section{Introduction}
\label{introduction}
Solar hard x-ray spectra in the deka-keV to 400 keV range are often best fit with a broken power-law \citep[e.g.,][]{1987ApJ...312..462L, 1992ApJ...389..756D, 2003ApJ...595L..97H} rather than a single power-law. Several mechanisms have been proposed to explain this knee-like spectral shape, including return-current losses. However, spectral fitting of the data with a return-current model is rare.\\
 
 Several studies have been devoted to interpreting the physical mechanisms responsible for producing a break in the electron and/or photon spectra. Such mechanisms include: (1) Non-uniform ionization along the path of electrons from the acceleration region into the thick target where they lose their energy by Coulomb collisions and emit hard x-rays, (hereafter HXRs) \citep[][]{1973SoPh...28..151B, 2003ApJ...595L.123K, 2009ApJ...705.1584S, 2011ApJ...731..106S}, (2) Compton back-scattering resulting in photospheric albedo \citep[][]{1978ApJ...219..705B, 2006A&A...446.1157K, 2007A&A...466..705K, 2011A&A...536A..93J},  (3) an anisotropic pitch angle distribution \citep[][]{1973ApJ...186..291P, 2004ApJ...613.1233M}, (4) beam-plasma instabilities \citep[][]{1982ApJ...257..354H, 1990SoPh..130....3M, 2009ApJ...707L..45H, 1977ApJ...218..866H, 2011A&A...529A.109H}, (5) return-current losses \citep[e.g.,][]{1977ApJ...218..306K,1982ApJ...257..354H} or (6) some process related to the electron accelaration mechanism \citep[e.g.,][]{1971SoPh...17..412L}, i.e., the injected electron distribution is different from a power-law, with a break or low-energy cutoff at energies above a few tens of keV.
It has been shown, in the above mentioned papers, that it is possible to rule out some of these mechanisms. This subject will be discussed further in section \ref{flare selection}.\\

Co-spatial return currents have been proposed to balance the electron flux (electrons s$^{-1}$) required to explain the observed x-ray bremsstrahlung emission  \citep[][]{1976SoPh...48..197H}. This return current \citep[][]{1970PhFl...13.1831H, 1971PhFl...14.1213L}, locally neutralizes the charge built up and cancels the magnetic field induced by the accelerated electron beam. \\    
     
The theory for co-spatial return currents has been developed in multiple papers. \cite{1977ApJ...218..306K} studied a classical return-current model in which an injected beam of electrons creates an electric field which in turn accelerates a stable return current. \cite{1980ApJ...235.1055E} and \cite{1981ApJ...249..817E} calculated analytical solutions for energy losses and heating rates from return currents and Coulomb collisions in a partially ionized atmosphere. \cite{1985A&A...142..219R} studied the conditions under which the return current becomes unstable to the generation of ion-acoustic or ion-cyclotron waves. Effects of a return current on x-ray emission have been investigated by \cite{1988SoPh..116..119D},\citeyearpar{1991SoPh..133..407D}, and \cite{1991SoPh..131..319L}. \cite{2006ApJ...651..553Z} numerically integrated the time-dependent Fokker-Planck equation to obtain the self-induced electric field strength and electron distribution function. \cite{2012ApJ...745...52H} obtained analytical solutions to return current losses in a fully ionized atmosphere, along with heating rates, electron distributions and x-ray brightness spectra in the context of a 1D model, which constitutes the basis for this paper.\\

Previously reported evidence for co-spatial return currents include \cite{2008A&A...487..337B}, who observed a spectral index difference between the loop top source and the footpoints of more than 2. This is more than the predicted difference in the standard thin/thick-target model. The authors conjectured that return-current losses might be responsible.  \cite{2007ApJ...666.1268A} found that the HXR flux from 10 flares saturates at a value 10$^{-15}$ photons cm$^{-2}$ s$^{-1}$ cm$^{-2}$, which corresponds to a limit predicted by \cite{1980ApJ...235.1055E}. \cite{2014AstL...40..499G} investigated the thermal runaway population of electrons from a hot plasma of temperature 100 MK, fitting a return current model to a white-light flare also obeserved by RHESSI and for which the electron flux density is high enough to suspect a significant effect from return current losses. However, there is no evidence for plasma temperatures of 100 MK in solar flares, and the fitted spectrum used in the paper is well-fitted with a single power-law distribution. In other words, since the photon spectrum does not show a flattening at lower energies, it is not possible to discriminate between the different models that could fit the spectra.  \cite{2013ApJ...773..121C} compared energy losses from four models. They studied these effects on electron energies between 13 and 30 keV. Note that this enegy range is likely to be dominated by thermal emission in large solar flares.\\

Beyond testing whether the \cite{2012ApJ...745...52H} return-current model can provide an acceptable fit to x-ray spectra, we are interested in plasma parameters and their time evolution that can be deduced when the model is acceptable. In particular, we test the return current stability to the generation of turbulence. For example, the resistivity of the coronal plasma can be deduced from the potential drop and injected electron flux density derived from the spectral fits. If the resistivity of the plasma is much higher than the classical Spitzer resistivity \citep[][]{1953PhRv...89..977S,1962pfig.book.....S} , then turbulence must be taken into account. It is not the subject of this paper to investigate turbulence mechanisms in detail \citep[][and references therein]{1985A&A...142..219R}; rather, we are interested in whether enhanced resistivity is necessary to explain the data (spectra and heating signatures), and whether the return current is stable. \cite{1985ApJ...293..584H} studied the effect of a runaway population of electrons and derived thresholds for producing ion acoustic, electrostatic ion cyclotron, and Buneman instabilities. Following \cite{2008A&A...487..337B}, who concluded that the return current would be unstable to wave growth for one of the two flares they studied, \cite{2013A&A...550A..63X} suggested that the return current could be unstable to ion acoustic instability resulting in enhanced (anomalous) resistivity. In this paper we derive a coronal resistivity from the parameters of the spectral fits to the measured x-ray photon spectra using Ohm's law, of which the validity is tested.
 \newline  
  
The model is introduced in section~\ref{model}. In section~\ref{method}  the method of analysis is explained. In section~\ref{sec:time evo}, two flares, 2005-Jan-19 and -20, are analyzed in detail. These serve as examples of the general analysis procedures and assessment of whether the return current model  can explain the observations. Section~\ref{stat} provides statistical results from 18 flares with acceptable return-current collisional thick-target model (RCCTTM) fits. Section~\ref{section5} explains the calculation method for coronal and beam/return current parameters. Section~\ref{stability} discusses the stability of the return current. Section~\ref{sec:heating} provides a discussion of the heating due to the return-current potential drop. Section~\ref{summary} summarizes the main results.\\

\section{The Model}
\label{model}
\subsection{Return-current collisional thick-target model (RCCTTM)}
\label{rccttm}
The standard thick-target model \citep[e.g.,][]{1971SoPh...18..489B,1972SvA....16..273S,  2011SSRv..159..107H} assumes that electrons are injected at the top of a flare loop and then stream toward the chromosphere. Along their path, the electrons are not subject to significant energy losses until they reach the thick-target where the density is high enough for Coulomb scattering and collisional losses to dominate. Consequently, the electron distribution is not significantly changed between the acceleration region and the footpoints. Other assumptions include that electrons are streaming in a cold, fully ionized atmosphere. In contrast, our model assumes that electrons lose some of their energy by return-current losses along their path toward the thick target, where they lose all of their remaining energy by Coulomb collisions. A cartoon representation of this model is shown in figure~\ref{fig:cartoon}.
Other assumptions are as follows:
\begin{figure}
\centering
\includegraphics[width=8.5cm]{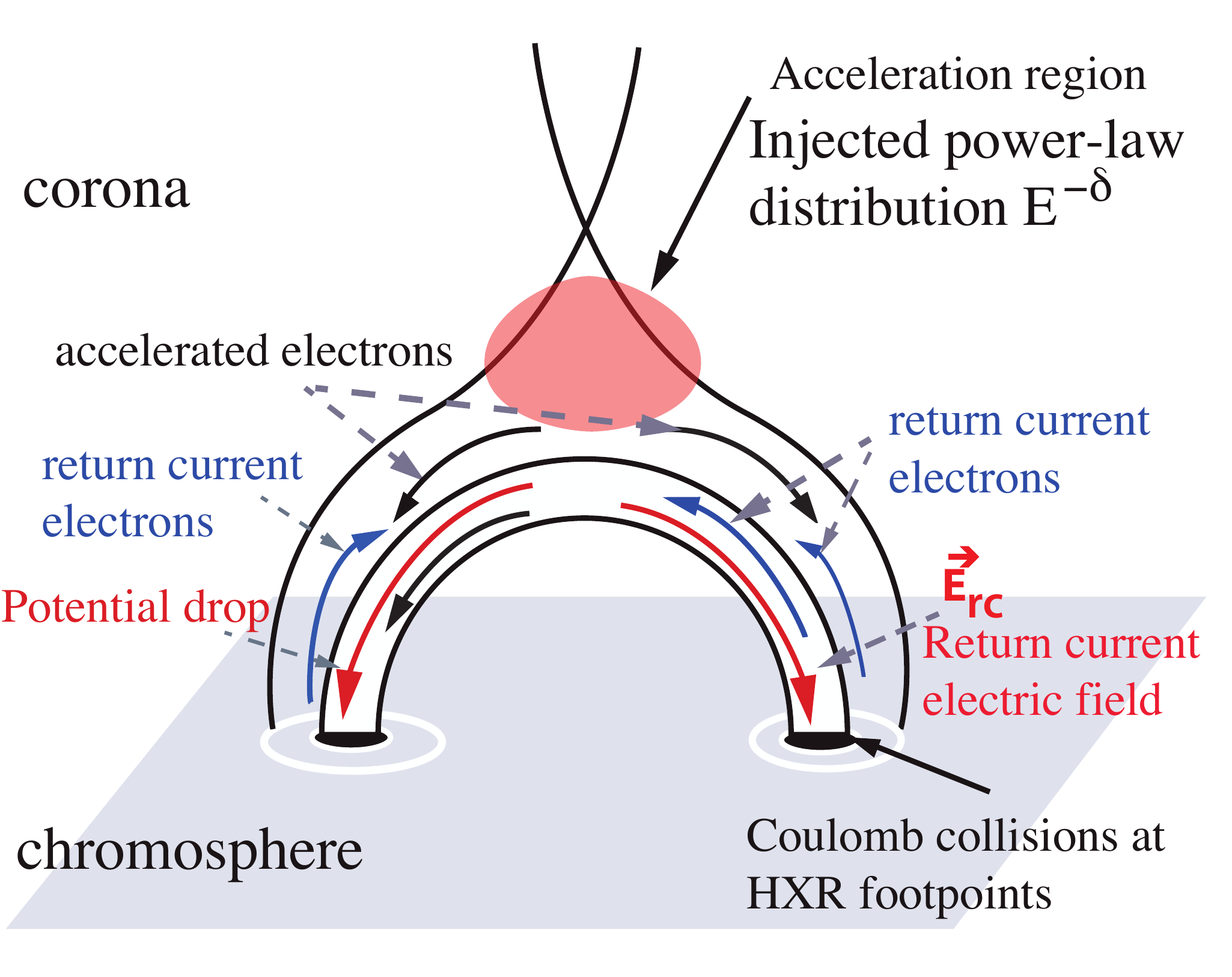}
\caption{Cartoon of the return-current collisional thick-target model. Electrons lose their energy due to the return-current electric field along the loop before they reach the chromosphere, where the density is high enough for significant collisional losses.}
\label{fig:cartoon}
\end{figure}

\begin{itemize}

 \item{Injected single power-law electron flux density energy distribution, with a sharp low-energy cutoff E$_{c0}$. 
 \begin{equation}F(E_0)=(\delta -1) E_{c0}^{(\delta-1)}F_{e0} E_0^{-\delta}\,\,\,\,\,\,\,\, [electrons\,\,\, s^{-1} \,\, cm^{-2} \,keV^{-1}] \label{eq1} \end{equation} for E$_{c0} <$ E$_0$,  where F$_{e0}$ is the injected electron flux density, $\delta$ is the electron spectral index at the acceleration region, and E$_0$ is the energy of an electron at the injection site.}
\item{The evolved electron distribution due to return-current losses is given by \cite{2012ApJ...745...52H}: 
 \begin{equation}{
F(E,x_{TT})=(\delta-1)  F_{e0}  {E_{c0}}^{\delta-1}  (E+e\,V_{TT})^{-\delta} 
\label{(2)}}
 \end{equation} 
  
where x$_{TT}$ is the distance from injection at the acceleration region to the thick target, e is the electron charge, V$_{TT}$ is the value of the potential drop at the thick target (x=x$_{TT}$), $E_{c0} = E_{c \,\,TT} + eV_{TT}$ is the injected low-energy cutoff, where E$_{c\,\,TT}$ is the value of the low-energy cutoff at x=x$_{TT}$. This condition is only valid when $E_{c\, TT}\,>\, E_{th}$.}
\item{Steady-state return-current density. This means that the current/return-current system reaches a steady state within the 8-s integration time for the fits. The time to reach the steady-state is on the order of the thermal electron-ion collision time, which is always less than 8 s for a plasma temeprature less than 80 MK \citep[][]{1990A&A...234..496V,Huba...2009}.}
\item{1D model with electrons streaming along the loop and co-spatial return current in the opposite direction.}
\item{Return-current losses dominate throughout the loop until the electrons reach the thick-target, where they lose all their energy by Coulomb collisions.}
\item{All the spectral flattening is due to return-current losses.}
\end{itemize}

The photon spectrum is the convolution of the electron distribution function F(E,x) with the relativistic electron-ion bremsstrahlung cross-section Q$(\epsilon, E)$, where $\epsilon$ and E are the photon and electron energies, respectively. Forward-fitting is used to fit RHESSI spectra with the RCCTTM. \\

Assumptions concerning the low-energy cutoff are discussed in the next section. The fit parameters are summerized in table~\ref{tab:model}. The relationship between F$_{50}$ and F$_{e0}$ is determined later in equations ~\ref{eq:fe0} and ~\ref{eq:fe0eth}. 

\begin{table}
\centering
\begin{tabular}{ c l  }
 \hline
& Nonthermal parameters\\
\hline
V$_{TT}$ & Return-current potential drop in kV at the thick-target\\
$\delta$ & Spectral index of the injected single power-law electron distribution\\
E$_{c\, \, TT\, \, \,\, max}$ & Maximum acceptable low-energy cutoff in keV at the footpoints\\
E$_{high}$ &  Injected high-energy cutoff in keV at the footpoints. This value is fixed at 32 MeV,\\
&  which is too high to have an effect on the spectral fits.\\
F$_{50}$ & Electron flux at  50 keV in electrons s$^{-1}$ keV$^{-1}$\\
\hline
& Thermal parameters\\
\hline
T & Temperature in keV\\
EM & Emission measure in cm$^{-3}$\\
\hline
\end{tabular}
\label{tab:model}
\caption{Fit parameters of the return current thick target model and an isothermal component.}
\end{table}

\subsection{Constraints on the low-energy cutoff}
\label{constraints}
The largest uncertainty is usually in determining the value of the low-energy cutoff, because the thermal population of electrons produces a steep bremsstrahlung spectrum at low energies which typically hides the low-energy cutoff \citep[][]{2011SSRv..159..107H, 2013ApJ...769...89I}. However, the range of possible energies for the low-energy cutoff is limited.\\

In the cold plasma approximation, electrons lose all of their energy in the thick target. However, since beam electrons propagate through a warm plasma, they can lose enough energy to be thermalized, at which point they are lost from the beam to the thermal background. \cite{2003ApJ...595L.119E} showed that in a warm target this thermalization energy is $E_{th}=\beta kT$ with $\beta \in [1,5]$. \cite{2015ApJ...809...35K} showed that if energy diffusion in a warm target is included, this lower limit should be approximately  E$_{th}$=$\delta$ kT, or greater. This provides a theoretical lower bound for the low-energy cutoff $E_{c\, min}=E_{th}$. We will assume that the loop temperature is constant in the corona and equal to the temperature derived from fitting the measured RHESSI x-ray spectrum to a single temperature model. This lower limit presents the extreme case where electrons are constantly thermalized from the looptop all the way to the thick target. As a consequence, the electron flux and return current electric field strength decrease as a function of the distance from the looptop.
 \\
 
It is possible in most cases to fit the nonthermal spectra with either a single or a double power-law electron distribution. However, the single and double power-law models do not provide information about the physics responsible for the shape of the spectrum. It is assumed that the fitted electron distribution entering the thick-target footpoints is unchanged from that at the injection site, at the loop top. In this paper the flattening at lower energies is explained by the potential drop associated with the electric field driving the return current. If the low-energy cutoff is too high, it will produce a spectral flattening inconsistent with the observed x-ray spectrum.\\

Since both a potential drop and a low-energy cutoff can produce almost the same spectral shape, we start by assuming a low-energy cutoff E$_c=$ 2 keV and fit the spectra with the return-current potential drop. This assumes that the flattening is entirely due to the return-current losses and not a combination of the low-energy cutoff and the potential drop. Then we increase the low-energy cutoff from 2 keV to the highest value that does not change the total $\chi^2$ by more than 1. This provides an upper limit on the value of the low-energy cutoff consistent with spectral flattening due to the potential drop. Note that this is usually higher than the maximum cutoff energy deduced from a single power-law fit and could be as high as the value of the energy loss due to the potential drop, but not higher. \\

 The upper limit to the low-energy cutoff corresponds to the case where all nonthermal electrons in the beam reach the thick target without being thermalized. This follows from the condition  $E_{c0\, max} -e\, V_{TT} = E_{c\,TT}\,\, > E_{th}$.

\section{Method of analysis}
\label{method}
\subsection{Flare selection}
\label{flare selection}

Significant progress has been made with RHESSI toward understanding the shape of solar x-ray spectra. All of the physical mechanisms described in this section can create a break in the spectrum. Here we explain how we chose flares with spectral breaks for which non-uniform ionization, albedo and anisotropy effects are insufficient to explain the observed breaks in the spectra.\\ 

Photons back-scattered from the photosphere (albedo) can contribute as much as 30$\%$ , if the emission is isotropic, to the photon spectrum at energies below 40 keV, adding a bump to the spectrum which could be interpreted as a break \citep[][]{2006A&A...446.1157K, 2007A&A...466..705K}. This bump is minimal at the solar limb. Most of the flares in our sample are limb events (15/19 flares) and hence have minimal contribution from Compton back-scattering. The four non-limb events were chosen because the spectral breaks were too large and/or the break enegy too high to be explained by albedo as shown in Table~\ref{mahtable}. We fit an albedo component from isotropically emitted photons to the spectra regardless of their position on the solar surface, using the \textit{albedo} routine in OSPEX.  \href{https://hesperia.gsfc.nasa.gov/ssw/packages/spex/doc/ospex_explanation.htm}{OSPEX} (Object Spectral Executive) is an object-oriented interface for x-ray spectral analysis of solar data.\\

\begin{table}
\centering
\begin{tabular}{ c c c c c c c c c c }
                         
 &  Flare & Peak  & Location & $\Delta \gamma$ & $\delta_{high}$ & E$_{br}$ & $\chi^2$ & Bkg & Reason for   \\
 &           & time  &                &                              &                           & [keV]        &                &  level &  selection  \\
        \hline

1   &   \href{https://hesperia.gsfc.nasa.gov/collaborate/malaouia/public_html/movies4html/20020226/movie.html}{2002-Feb-26}   &  10:26    &   [927,-228]    & 0.48    & 4.34 & 76 &1.21,   0.98 &16  &    Sui/Su  \\

2  &    \href{https://hesperia.gsfc.nasa.gov/collaborate/malaouia/public_html/movies4html/20020602/movie.html}{2002-Jun-02}   &   11:46    &   [-152,-305]   & 1.54    & 4.83 & 50* &0.8,  0.97 &2.6  &   Sui \\               

3  &   \href{https://hesperia.gsfc.nasa.gov/collaborate/malaouia/public_html/movies4html/20020822/movie.html}{2002-Aug-22}   &   01:52    &   [826,-273]    & 1.16    & 5.42   &  83  & 1.11,  0.71  & 3.8  & SB \\

4  &   \href{https://hesperia.gsfc.nasa.gov/collaborate/malaouia/public_html/movies4html/20020828/movie.html}{2002-Aug-28}   &   10:59    &   [884,-314]    & 1.17    & 4.59 & 59* &0.89,  0.83 & 3.2 &    Su   \\       
5  &   \href{https://hesperia.gsfc.nasa.gov/collaborate/malaouia/public_html/movies4html/20030608/movie.html}{2003-Jun-08}    &   16:09   &   [-784,-304]  & 0.88    & 5.58   & 119   & 1.04,  1.03  & 5.5  & SB \\

6  &   \href{https://hesperia.gsfc.nasa.gov/collaborate/malaouia/public_html/movies4html/20030613/movie.html}{2003-Jun-13}    &   04:34    &   [933,172]    & 0.85    & 5.04 & 77 & 0.68,  0.45 & 1.7  &   Su  \\
7  &   \href{https://hesperia.gsfc.nasa.gov/collaborate/malaouia/public_html/movies4html/20030617/movie.html}{2003-Jun-17}    &   22:46    &   [-813,-142]  &  0.68   & 4.51 &189 & 1.11,  1.36 & 12 & SB   \\
8  &  \href{https://hesperia.gsfc.nasa.gov/collaborate/malaouia/public_html/movies4html/20031103/movie.html}{2003-Nov-03}  &   09:49   &   [929,135]     & 0.59    & 5.20 &137&0.59,  0.70 & 48 & SB \\
9 &   \href{https://hesperia.gsfc.nasa.gov/collaborate/malaouia/public_html/movies4html/20050117/movie.html}{2005-Jan-17}   &   09:42 & [356,317]  & 0.75     &  4.86 & 72 & 0.70,  0.69 & 2.5 & SB \\
10  &  \href{https://hesperia.gsfc.nasa.gov/collaborate/malaouia/public_html/movies4html/20050119/movie.html}{2005-Jan-19}   &   08:25   &   [739,313]     & 0.67    & 3.92 & 190 & 0.79,  0.79 & 74  & Warmuth \\
11  &  \href{https://hesperia.gsfc.nasa.gov/collaborate/malaouia/public_html/movies4html/20050120/movie.html}{2005-Jan-20}   &   06:45   &   [853,276]     & 0.51    &3.75 & 134 & 1.07,  0.89 & 273&  SB     \\
12  &  \href{https://hesperia.gsfc.nasa.gov/collaborate/malaouia/public_html/movies4html/20050121/movie.html}{2005-Jan-21}   &   10:14  &    [896,322]     & 0.81    &5.01 & 107 & 1.18,  0.74 & 3.1 &  SB     \\
13  &  \href{https://hesperia.gsfc.nasa.gov/collaborate/malaouia/public_html/movies4html/20050730/movie.html}{2005-Jul-30}   &   06:32   &    [-811,133]    &  0.72   & 4.40 & 106 & 1.03,  1.12 &  2.1  & SB\\
14  &  \href{https://hesperia.gsfc.nasa.gov/collaborate/malaouia/public_html/movies4html/20061206/movie.html}{2006-Dec-06 } &   18:43   &    [-855,-117]   & 0.31    & 3.30  & 107* & 0.40,  0.60 & 203 & SB \\
15  &  \href{https://hesperia.gsfc.nasa.gov/collaborate/malaouia/public_html/movies4html/20061214/movie.html}{2006-Dec-14}  &    22:09   &  [690,-93]    & 0.75    & 5.10 & 80   &0.78,  0.70  & 5.9 &  SB  \\
16  &  \href{https://hesperia.gsfc.nasa.gov/collaborate/malaouia/public_html/movies4html/20110926/movie.html}{2011-Sep-26}  &   05:05   &    [-522,120]   & 0.89     &  4.19 & 47*   &0.98,  1.07   & 13 &  SB/EI  \\
17  &  \href{https://hesperia.gsfc.nasa.gov/collaborate/malaouia/public_html/movies4html/20130513/movie.html}{2013-May-13}  &   16:03  &    [-949,177]   &  0.54    & 3.51 &123 &  0.97,  1.17 &  56 & SB      \\
18  &  \href{https://hesperia.gsfc.nasa.gov/collaborate/malaouia/public_html/movies4html/20140329/movie.html}{2014-Mar-29}   &  17:45  &    [514,267]    &  0.65    & 3.54 & 49* & 0.76,  0.81   &  11.26  &  SB \\
19  &  \href{https://hesperia.gsfc.nasa.gov/collaborate/malaouia/public_html/movies4html/20150505/movie.html}{2015-May-05}  &   22:14  &    [-904,250]    & 0.48    &  3.94& 164  & 1.16,  1.10  & 59  &  SB/EI \\

 \hline  \hline
\end{tabular}
\caption{Event list and spectral fit parameters at the time of peak emission. $\Delta\gamma= \gamma_{high}- \gamma_{low}$ is the spectral index difference from the broken power-law fits to the photon spectra, $\delta_{high}$ is the high energy electron spectral index using the standard collisional thick-target model (CTTM) with a broken power-law electron distribution and sharp low-energy cutoff, E$_{br}$ is the break energy from the electron double power-law fits. The asterisks in the E$_{br}$ column mean the spectrum was better fitted with a single power-law than a double power-law and the energy in that column is the low-energy cutoff that is consistent with the break in the photon spectrum; the two reduced $\chi^2$ values are from the photon and electron fits respectively, the background level is the ratio of the source photon to the background spectra at 105 keV, and the last column gives the reason for chosing the flare: \textit{Sui} for early impulsive flares from \cite{2007ApJ...670..862S}, \textit{Su} for flares that cannot be explained by non-unifrom ionization from \cite{2009ApJ...705.1584S} , \textit{SB} for flares with strong breaks, \textit{EI} for early impulsive flares, and  \textit{Warmuth} for the flare with a high low-energy cutoff at 120 keV published by \cite{2009ApJ...699..917W}.  Note that the date column contains a link to a movie of the spectral fits for each flare.}
\label{mahtable}
\end{table}

 Non-uniform ionization can result in a spectral break as high as 0.6, with a higher power-law index in the electron distribution corresponding to a higher spectral index difference in the photon spectrum \citep[][]{1973SoPh...32..227B, 2009ApJ...705.1584S}. We choose flares for which the spectral index difference is higher than what is expected from non-uniform ionization alone \citep[][]{2009ApJ...705.1584S}. \\  
 
 We correct for possible pulse pile-up by adding a pile-up correction component while fitting the spectra, using the default parameters in OSPEX.\\ 
 
We focus on early impulsive flares \citep[e.g.,][]{2007ApJ...670..862S} and flares with strong spectral breaks (labeled SB in Table ~\ref{mahtable}). Early impulsive (EI) flares have the advantage of observing the emission from the nonthermal population of electrons at low x-ray energies, before significant thermal bremsstrahlung dominates this emission. If these EI flares were studied in \cite{2007ApJ...670..862S}, we refer to them as Sui in Table~\ref{mahtable}. The resulting sample of flares is given in Table ~\ref{mahtable}. The 2005-Jan-19 flare was studied by \cite{2009ApJ...699..917W}, who found the flattening after 08:24 UT is due to a value of the cutoff energy $\sim$120 keV. We refer to the flares that have a break higher than can be explained by non-uniform ionization, studied by \cite{2009ApJ...705.1584S}, as Su. \\

We fit spectra between 12 and $\leq$300 keV. Fitting the spectra at energies $\geq$12 keV in large flares ensures the fitted temperature is the highest temperature in the loop. There is usually no improvement of the determination of this temperature if we fit the data down to 6 keV, because of the complexity of the spectrum in the energy range between 6 and 12 keV, which contains an Fe line complex at 6.7 keV, the Fe/Ni line complex  at 8 keV, plus an instrumental line at $\sim$10 keV. Fitting the spectra to $\leq$300 keV ensures that electron-electron bremsstrahlung is insignificant \citep[][]{2007ApJ...670..857K}, and allows the use of only the front detector segments.\\

\subsection{Source sizes}
\label{source sizes}
Footpoint areas are necessary to compute the electron flux density, which is a fundamental quantity in the return-current model, as the return-current electric field is proportional to it through Ohm's law. The length of the loop is also necessary to compute the return-current electric field from the potential drop. These will be discussed in detail in section~\ref{section5}. 

The area of the footpoint source is estimated by using CLEAN \citep[][]{1974A&AS...15..417H} images integrated over the duration of the hard x-ray peak time. Ideally, the time integration of the reconstructed images should match the time integration of the spectra, which is 4 s or 8 s in this study. Unfortunately, the count rate is usually insufficient to reliably separate the signal from the noise on such short times. We had to integrate over longer time periods,  shown in column 3 of Table~\ref{tab:sourcestable}. The values obtained are then upper limits on the footpoint areas.\\

Decisions regarding energy bands, detectors used and time of integration to reconstruct the images are summarized in Table ~\ref{tab:sourcestable}. The same integration times were used to reconstruct both the thermal sources at 12-20 keV and the nonthermal sources at higher energies given in column 7. The footpoint areas shown in column 4 of table~\ref{tab:sourcestable} are then determined by summing within the 50\% contour of each footpoint. Context images from instruments including TRACE, EIT and SDO AIA were used to match the x-ray sources with EUV ribbons when possbile. \\

The loop half-length is estimated as follows: $h_{loop} ={{\pi R } \over 2}$ where 2R is the distance between the footpoint source centroids. During the 17-Jun-2003, the only flare in our sample where the footpoint source is not resolved into multiple footpoints, we take the area of one footpoint to be the total area divided by 2, and the distance between the footpoints 2R is estimated as R$ = {1 \over 2 }{\sqrt{{A_{total}} \over {\pi}}}$. When the nonthermal source has three footpoints (3/19 events), we use the EUV context images when available to determine which two footpoints align with the same ribbon, average the distance between the 2 centroids, use the new value as the centroid of one footpoint, and then calculate the loop half-length, similarly to the two-footpoint case. \\

The thermal source volume $\Omega$ is used to estimate the background density. Since the flare spectra in our sample are fitted above 12 keV, the thermal source is integrated between 12 and 20 keV. This energy band is dominated by thermal emission in the sample of flares presented. This volume is calculated by taking the 50$\%$ contour area to the power of $3/2$: $\Omega = A_{12-20keV}^{3/2}$\\

The estimated values in Table~\ref{tab:sourcestable} account for results from \cite{2009ApJ...698.2131D}. The authors compared different image-reconstruction algorithms in order to constrain compact source sizes. In particular, they showed that the default CLEAN beam width factor, which is a Gaussian with a width equal to the width of the point spread function, can lead to over-estimation of the source size. A better estimate of the source size is obtained by doubling the beam width factor, which actually means that the FWHM of the new Gaussian is divided by 2. We set the beam width factor to 2 when detectors 3 and above were used, and the default value equal to 1 otherwise.   \\

\begin{table}
\centering
\begin{tabular}{ c c c c c c c  }
                         
Flare $\#$ & Date & Time    &    Footpoint  & Loop  &Detectors & Energy  \\
  &        &   range &    area  & half-length      & & [ keV ]     \\
  &        &    &    $\times$10$^{16}$ [cm$^{2}$] &  [ Mm ]     & &     \\
  
\hline
1 &\href{https://hesperia.gsfc.nasa.gov/collaborate/malaouia/public_html/movie4html/source20020226.pdf}{2002-Feb-26} & 10:26:16+80s &  7.4    & 20    & 1,2,3,4,5,6,7&  40-95  \\

2 &\href{https://hesperia.gsfc.nasa.gov/collaborate/malaouia/public_html/movie4html/source20020602.pdf}{2002-Jun-02}  & 11:44:24+36s  & 26   & 7.6   & 3,4,5,6,7,8,9&  35-95   \\
3 &\href{https://hesperia.gsfc.nasa.gov/collaborate/malaouia/public_html/movie4html/source20020822.pdf}{2002-Aug-22}*&     01:52:00+60s   & 24    & 29   &2,4,5,6,7,8 &  50-100    \\

4 &\href{https://hesperia.gsfc.nasa.gov/collaborate/malaouia/public_html/movie4html/source20020828.pdf}{2002-Aug-28}  &     10:59:08+36s   & 4.4    & 5.9  & 1,2,4,5,6,7,8&  35-90     \\

5 &\href{https://hesperia.gsfc.nasa.gov/collaborate/malaouia/public_html/movie4html/source20030608.pdf}{2003-Jun-08}  &   16:09:20+24s  & 26   & 6.8-14   & 2,3,4,5,6,7,8 &  50-100    \\

6 &\href{https://hesperia.gsfc.nasa.gov/collaborate/malaouia/public_html/movie4html/source20030613.pdf}{2003-Jun-13} &   04:34:08+36s  &  5.9    &7.3  & 1,2,3,4,5,6,7,8&  40-90  \\
7 &\href{https://hesperia.gsfc.nasa.gov/collaborate/malaouia/public_html/movie4html/source20030617.pdf}{2003-Jun-17}  &     22:46:12+36s  & 14    & 14-27   & 2,3,4,5,6,7,8 &  60-130    \\
   & \href{https://hesperia.gsfc.nasa.gov/collaborate/malaouia/public_html/movie4html/source20030617.pdf}{2003-Jun-17}   &     22:48:32+36s  & 29    & 27   & 2,3,4,5,6,7,8 &  60-130    \\
    &\href{https://hesperia.gsfc.nasa.gov/collaborate/malaouia/public_html/movie4html/source20030617.pdf}{2003-Jun-17}   &     22:53:12+36s  & 27    & 13   & 2,3,4,5,6,7,8 &  60-130    \\

8 &\href{https://hesperia.gsfc.nasa.gov/collaborate/malaouia/public_html/movie4html/source20031103.pdf}{2003-Nov-03} &   09:49:06+64s  &  56  &   9.0& 3,4,5,6,7,8 &80-110\\
  &\href{https://hesperia.gsfc.nasa.gov/collaborate/malaouia/public_html/movie4html/source20031103.pdf}{2003-Nov-03}&   09:51:48+64s  &  33 &   12& 3,4,5,6,7,8 &80-110\\
 & \href{https://hesperia.gsfc.nasa.gov/collaborate/malaouia/public_html/movie4html/source20031103.pdf}{2003-Nov-03}&   09:56:56+64s  &  34  &   16& 3,4,5,6,7,8 &80-110\\

9 &\href{https://hesperia.gsfc.nasa.gov/collaborate/malaouia/public_html/movie4html/source20050117.pdf}{2005-Jan-17}  &  09:43:08+54s & 32    & 13-22  & 2,4,5,6,7,8 &  80-120  \\
  &\href{https://hesperia.gsfc.nasa.gov/collaborate/malaouia/public_html/movie4html/source20050117.pdf}{2005-Jan-17}   &  09:44:36+48s & 16    & 21  & 2,4,5,6,7,8 &  80-120  \\
  &\href{https://hesperia.gsfc.nasa.gov/collaborate/malaouia/public_html/movie4html/source20050117.pdf}{2005-Jan-17}   &  09:48:32+48s & 58  & 21-26  & 2,4,5,6,7,8 &  80-120  \\
 &\href{https://hesperia.gsfc.nasa.gov/collaborate/malaouia/public_html/movie4html/source20050117.pdf}{2005-Jan-17}  &  09:49:56+54s & 33   & 26   & 2,4,5,6,7,8 &  80-120  \\
  &\href{https://hesperia.gsfc.nasa.gov/collaborate/malaouia/public_html/movie4html/source20050117.pdf}{2005-Jan-17}   &  09:52:56+54s & 24  &  33  & 2,4,5,6,7,8 &  80-120  \\

10 &\href{https://hesperia.gsfc.nasa.gov/collaborate/malaouia/public_html/movie4html/source20050119.pdf}{2005-Jan-19}  &  08:12:22+68s & 37    & 18   & 3,4,5,6,7,8 &  80-120  \\
 &\href{https://hesperia.gsfc.nasa.gov/collaborate/malaouia/public_html/movie4html/source20050119.pdf}{2005-Jan-19}   &  08:15:10+76s & 44    & 24   & 3,4,5,6,7,8 &  80-120  \\
 &\href{https://hesperia.gsfc.nasa.gov/collaborate/malaouia/public_html/movie4html/source20050119.pdf}{2005-Jan-19}   &  08:16:58+72s & 21  & 28   & 3,4,5,6,7,8 &  80-120  \\
 &\href{https://hesperia.gsfc.nasa.gov/collaborate/malaouia/public_html/movie4html/source20050119.pdf}{2005-Jan-19}   &  08:19:42+132s & 23   & 32 & 3,4,5,6,7,8 &  80-120  \\
 &\href{https://hesperia.gsfc.nasa.gov/collaborate/malaouia/public_html/movie4html/source20050119.pdf}{2005-Jan-19}   &  08:25:00+120s & 26   & 37  & 3,4,5,6,7,8 &  80-120  \\

11 &\href{https://hesperia.gsfc.nasa.gov/collaborate/malaouia/public_html/movie4html/source20050120.pdf}{2005-Jan-20}  &  06:44:10+64s & 13    & 17  & 1,2,3,4,5,6,7,8 &  80-120  \\
 &\href{https://hesperia.gsfc.nasa.gov/collaborate/malaouia/public_html/movie4html/source20050120.pdf}{2005-Jan-20}   &  06:52:52+124s & 39   & 18   &2,3,4,5,6,7,8,9  &  60-100  \\

12 &\href{https://hesperia.gsfc.nasa.gov/collaborate/malaouia/public_html/movie4html/source20050121.pdf}{2005-Jan-21}   &   10:14:44+36s &  20    & 8.1   & 3,4,5,6,7,8 &  70-130 \\

13 &\href{https://hesperia.gsfc.nasa.gov/collaborate/malaouia/public_html/movie4html/source20050730.pdf}{2005-Jul-30}  &  06:31:00+96s  & 13    & 14    &2,3,4,5,6,7,8 &  50-120   \\
 & \href{https://hesperia.gsfc.nasa.gov/collaborate/malaouia/public_html/movie4html/source20050730.pdf}{2005-Jul-30} &  06:34:00+96s  & 13   & 15    &2,3,4,5,6,7,8 &  50-120  \\

14 &\href{https://hesperia.gsfc.nasa.gov/collaborate/malaouia/public_html/movie4html/source20061206.pdf}{2006-Dec-06} &  06:43:04+56s  & 26   & 11    & 1,4,6,7,8,9 &  50-120   \\
 &\href{https://hesperia.gsfc.nasa.gov/collaborate/malaouia/public_html/movie4html/source20061206.pdf}{2006-Dec-06} &  06:51:20+120s  & 6.4  & 30   & 1,4,6,7,8,9 &  50-120   \\
15 &\href{https://hesperia.gsfc.nasa.gov/collaborate/malaouia/public_html/movie4html/source20061214.pdf}{2006-Dec-14} &  22:08:28+64s  & 22    & 16    & 2,3,4,5,6,7,8 &  60-110   \\

16 &\href{https://hesperia.gsfc.nasa.gov/collaborate/malaouia/public_html/movie4html/source20110926.pdf}{2011-Sep-26} & 05:06:12+12s & 30 & 6.3  & 2,3,4,5,6,7,8,9 & 50-120 \\

17 &\href{https://hesperia.gsfc.nasa.gov/collaborate/malaouia/public_html/movie4html/source20130513.pdf}{2013-May-13}   &   16:00:42+64s &  13    & 12   &1,3,4,5,6,7,8 &  60-120 \\
 & \href{https://hesperia.gsfc.nasa.gov/collaborate/malaouia/public_html/movie4html/source20130513.pdf}{2013-May-13}    &   16:02:12+64s &  8.8    & 13   &1,3,4,5,6,7,8 &  60-120 \\

18 &\href{https://hesperia.gsfc.nasa.gov/collaborate/malaouia/public_html/movie4html/source20140329.pdf}{2014-Mar-29} & 17:46:10+82s &  18.3    & 9.7   & 1,2,3,4,5,7,8 &  50-110  \\

19 &\href{https://hesperia.gsfc.nasa.gov/collaborate/malaouia/public_html/movie4html/source20150505.pdf}{2015-May-05} & 22:07:48+64s  &  5.9  & 16  &1,3,5,7,8 &  60-120  \\
 &\href{https://hesperia.gsfc.nasa.gov/collaborate/malaouia/public_html/movie4html/source20150505.pdf}{2015-May-05}  & 22:09:24+90s  &  8.3    & 8.9   &1,3,5,7,8 &  60-120  \\

 \hline  \hline

\end{tabular}
\caption{RHESSI footpoint areas and loop half-lengths. The asterisk next to the 2002-08-22 date means it is the only flare not included in the statistical analysis of section~\ref{stat}. The time range column gives the integration interval start time and duration in seconds. The footpoint area is the total area of the footpoints deduced from the 50$\%$ contour levels. Columns 6 and 7 are the detectors and energy range used to calculate the footpoint areas, respectively. Note that the date column contains a link to the source images.}

\label{tab:sourcestable}
\end{table}

The integration time of each image is longer than the 8-s integration time of the fitted spectra, because more counts are needed to obtain a reliable source size. We obtain the highest resolution possible within the integration time interval by using all detectors that show a modulated signal.\\

\subsection{Uncertainties}
\label{uncertainties}
Uncertainties on spectral fit parameters are calculated using a Monte Carlo analysis  \citep[see section 3.1.3 in][]{2013ApJ...769...89I}. A sample is drawn from a Poisson distribution in each energy bin and a new count spectrum is constructed from the expected fitted count spectrum. Each count spectrum is fitted to obtain a new set of fit parameters. The sample and fit process is repeated 1000 times. No systematic uncertainty is included in our analysis.\\

Note that the uncertainty distribution of parameters can be asymmetric. The 68$\%$ and 95$\%$ probability intervals are calculated by removing 16$\%$ and 2.5$\%$ tails on each side of the distribution, respectively; these correspond to 1$\sigma$ and 2 $\sigma$ intervals, when the distribution is symmetric and normal. The probability intervals of other plasma parameters that are not directly fitted by the model, such as the return-current drift velocity or the injected flux density, are computed by generating a 1000 point distribution from the fitted parameters, then removing the 16$\%$ and 2.5$\%$ tails of the distribution.  \\

We run the Monte Carlo analysis 1000 times with the low-energy cutoff fixed at 2 keV and the potential drop free; then 1000 times with the potential drop fixed at the most probable value and the low-energy cutoff free. The first is to estimate the uncertainties on the potential drop under the assumption that the flattening is fully due to return-current losses, and the second is to estimate the maximum low-energy cutoff consistent with the fitted potential drop.  1000 runs is sufficient to estimate the 95$\%$ confidence level, as this provides 25 cases on each tail of the distribution to be removed. Note that we use the 67$\%$ confidence level throughout this paper. Te number of runs is limited to 1000 because this is a time-consuming process, as there are 1154 spectra in our sample and the Monte Carlo routine is run 2000 times for each spectrum.\\

\section{Time evolution results for 2005-Jan-19 -20 flares}
\label{sec:time evo}
\subsection{Spectral fits}
 \href{https://hesperia.gsfc.nasa.gov/collaborate/malaouia/public_html/movies4html/20050119/movie.html}{2005-Jan-19}{\footnote{Click on the date for a movie of the spectral fits}\label{footnote1}} is an X1.3 class flare that was studied by \cite{2009ApJ...699..917W}. The authors deduced that the last HXR peak at $\sim$08:24 UT is consistent with a value of the low-energy cutoff around 120 keV. They argue that a high value of the low-energy cutoff is compatible with the absence of the Neupert effect in this peak. Indeed, if there are few electrons with energies below $\sim$120 keV injected at the loop top, there will be no significant energy losses relative to the energy content of the flare plasma, and therefore no significant heating of the plasma. Here we fit the spectra from this event with the RCCTTM. In other words, we test whether return current losses are consistent with the spectral flattening observed at the lower energies.\\
  
\href{https://hesperia.gsfc.nasa.gov/collaborate/malaouia/public_html/movie4html/20050120/movie.html}{2005-Jan-20}$^{1}$ was chosen as another example because the derived potential drop values are amongst the highest in our sample.\\

\begin{figure}
\includegraphics[width=\textwidth]{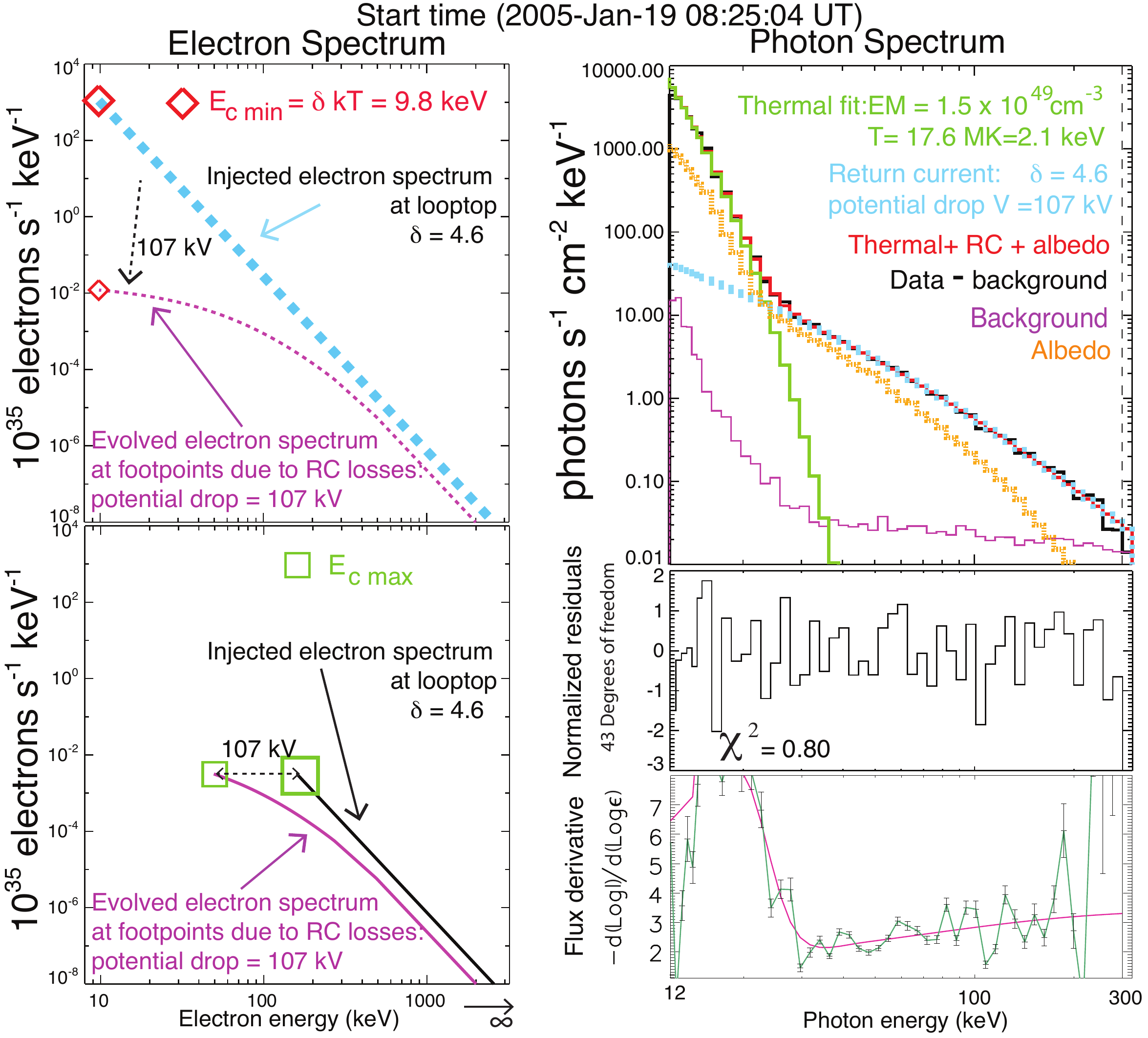}
\caption{Example of a model electron distribution and corresponding photon spectrum. \textit{Left panels:} Electron distribution model used to fit the photon spectrum. The black (blue dashed) curve is the injected electron distribution with the sharp low-energy cutoff at $\sim$152 keV (9.8 keV) equal to the maximum (minimum) fitted low-energy cutoff of 45 keV plus the fitted potential drop of 107 kV ($\delta$kT, where T is the fitted temperature and $\delta$ the fitted spectral index). The purple curve is the evolved electron distribution due to return current losses at the footpoints. This is the fitted electron model, whereas the black and blue spectra are deduced from the assumptions of our model. The green square is the highest fitted low-energy cutoff consistent with a potential drop of 107 kV, the red diamond is the lower limit on the low-energy cutoff $\delta$kT= 9.8 keV {\textit{Upper right panel:}} Photon spectrum fit. Black: RHESSI data with the background substracted; green: isothermal fit; light blue: return current model; red: sum of isothermal, return current model, and albedo component; orange: albedo component at position (741'', 316''); purple: background spectrum. \textit{Middle right panel:} normalized residuals with a reduced $\chi^2$ of 0.80; the number of degrees of freedom is 43. {\textit{Lower right panel:}} Derivative of the photon flux logarithm. The solid pink and green lines are the model and data, respectively.}
\label{fig:spectrum}
\end{figure}

Figure ~\ref{fig:spectrum} shows an example of a spectral fit using the return current model. This is a case where the spectral flattening in the photon spectrum is visually clear, and a relatively high potential drop of about 107 kV was deduced from the spectral fits. The left panels represent the electron model fit used for the lower limit of the low-energy cutoff in the upper panel and the upper limit in the lower panel. The purple curves give the fitted electron distribution at the footpoints with the upper and lower limits of the low-energy cutoff represented by a green square and a red diamond, respectively. The deduced injected electron distribution at the loop top is plotted in black and blue in the upper and lower limits on the cutoff energy, respectively. The green square (red diamond) along the black (blue) spectrum is the upper (lower) limit on the cutoff energy. \\

Since the fitted upper limit on the low-energy cutoff is 45 keV, with a potential drop between the looptop and the footpoints of 107 kV the upper limit on the cutoff energy at injection is 152 keV: $E_{c\, max} = E_{c\,\, TT}\, +e\, V_{TT}$. The lower limit of the low-energy cutoff is constant along the loop and equal to $\delta$ k T, where the temperature is taken to be constant along the loop and equal to the temperature determined from the isothermal fit.\\

The upper right panel of figure~\ref{fig:spectrum} shows the observed background-subtracted x-ray spectrum in black, with the background spectrum shown in purple. We fit energies between 12 keV and 280 keV. The upper limit is determined by the background-subtracted flux levels. This energy is defined  by the lower limit of the energy bin with 10 counts/bin. Therefore, the fitted energy range changes throughout the flare.  The lower limit is set at 12 keV. This is justified by most flares in our sample being high count rate flares, where both RHESSI attenuators are used to prevent pulse pile-up, with 12 keV being the lower energy limit where the count rate could be reliably determined. Another reason for chosing 12 keV as the lower limit of the fitted energy range is to fit the higher temperature component. \cite{2010ApJ...725L.161C} showed that RHESSI spectra could be better fitted with multiple thermal components in large flares.\\

The blue curve is the return-current fit with a spectral index of 4.6 and a potential drop of 107 kV, the green curve is the isothermal fit with a temperature of 17.6 MK and an emission measure of 1.5 $\times$ 10$^{49}$ cm$^{-3}$. The albedo component, assuming the emission is isotropic, is plotted in orange at the position of the flare, (741'', 316'') from disk center. Finally, the model photon spectrum which is the sum of the isothermal, and return-current models and the albedo component is represented by the red curve.\\ 

The energy binning used is the same for all flares. It was manually defined by 0.5 keV-sized bins between 12 and 15 keV and then 45 logarithmically-spaced bins between 15 and 300 keV with 51 energy bins in total between 12 and 300 keV.  The lower right panel is the normalized residuals with a reduced $\chi^2$ value of 0.8.

  \subsection{Time evolution}
 \label{sec:timeevo} 
\begin{figure*}[t]
  \begin{minipage}[c]{.62\linewidth}
    \null
\includegraphics[width=\textwidth]{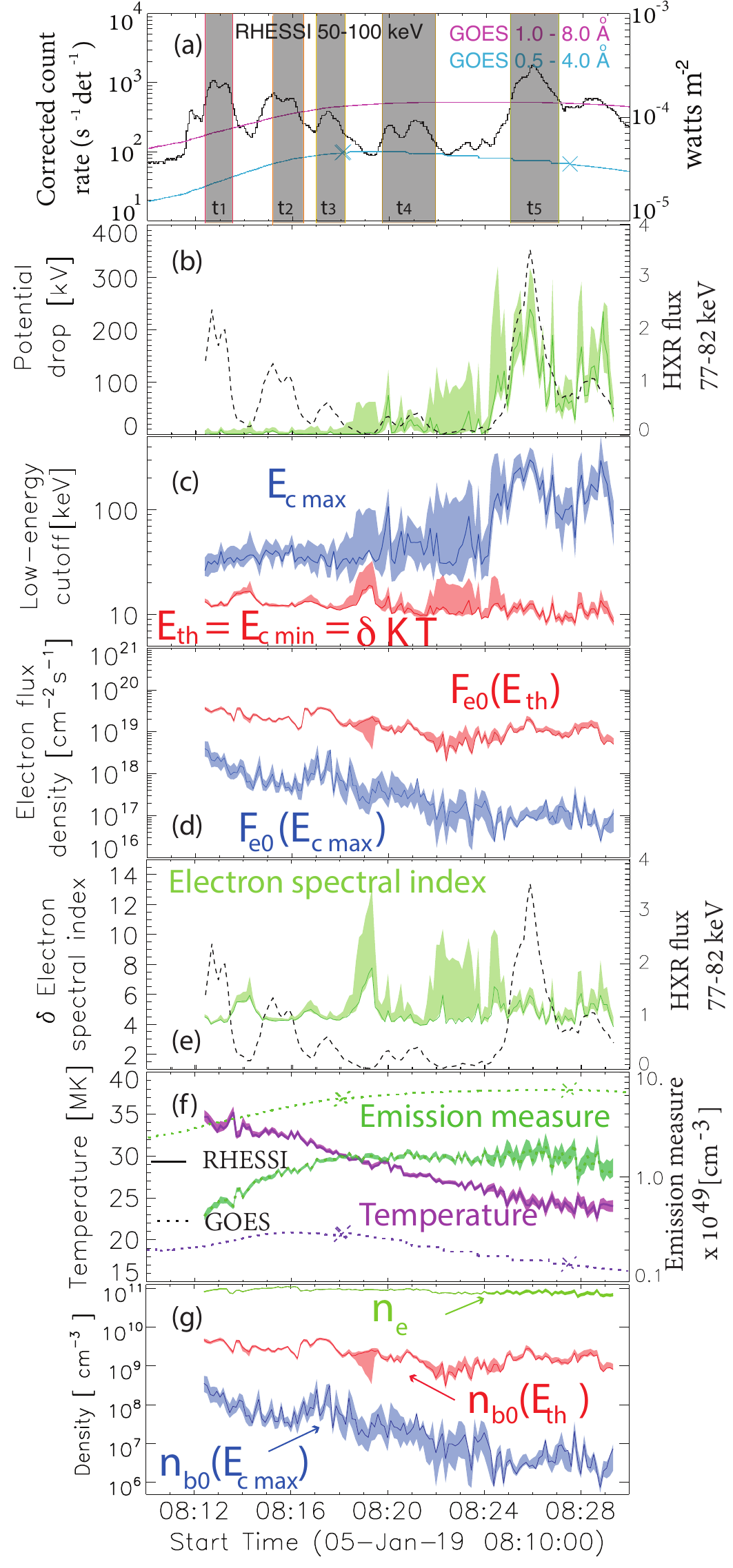}
\end{minipage}\hfill
 \begin{minipage}[c]{0.32\textwidth}
\caption{ Time evolution of fit parameters from 2005-Jan-19. The width of all curves in a lighter color represent the 67$\%$ confidence interval from Monte Carlo runs and the curves in the darker color are the modes. {\textit{(a):}} The left axis represents the RHESSI HXR lightcurve integrated between 50 and 100 keV in counts s$^{-1}$ detector$^{-1}$, and the right axis is the GOES flux in W/m$^2$, with the pink/blue curves being GOES lightcurves at two wavelength bands: 1-8 $\AA$ and 0.5-4 $\AA$, respectively. The grey shaded areas represent the integration times of the images of the  \href{https://hesperia.gsfc.nasa.gov/collaborate/malaouia/public_html/movie4html/source20050119.pdf}{source sizes} which can be found online; {\textit{(b):}} Time evolution of the potential drop in kV on the left axis, HXR flux in the nonthermal energy bin 77-82 keV;  {\bf\textit{(c):}} The red/blue curves show the time evolution of the low-energy cutoff at the looptop in the lower/upper limits respectively; {\textit{(d):}} Electron flux density at the looptop in upper/lower limits of low-energy cutoff in blue/red respectively. Note that the same color code is used for (c), (d) and (g) {\bf{\textit{(e):}}} Electron spectral index in green and HXR flux in black  {\textit{(f):}} Temperature in purple (left axis) and emission measure in green (right axis). RHESSI and GOES data are the solid and dashed curves, respectively. {\textit{(g):}} Density of the thermal background in green, the beam density at the looptop in the upper/lower limits of the cutoff energy (blue/red).}
\label{fig:timeevo}
 \end{minipage}
 
\end{figure*}

\begin{figure*}[t]
  \begin{minipage}[c]{.62\linewidth}
    \null
\includegraphics[width=\textwidth]{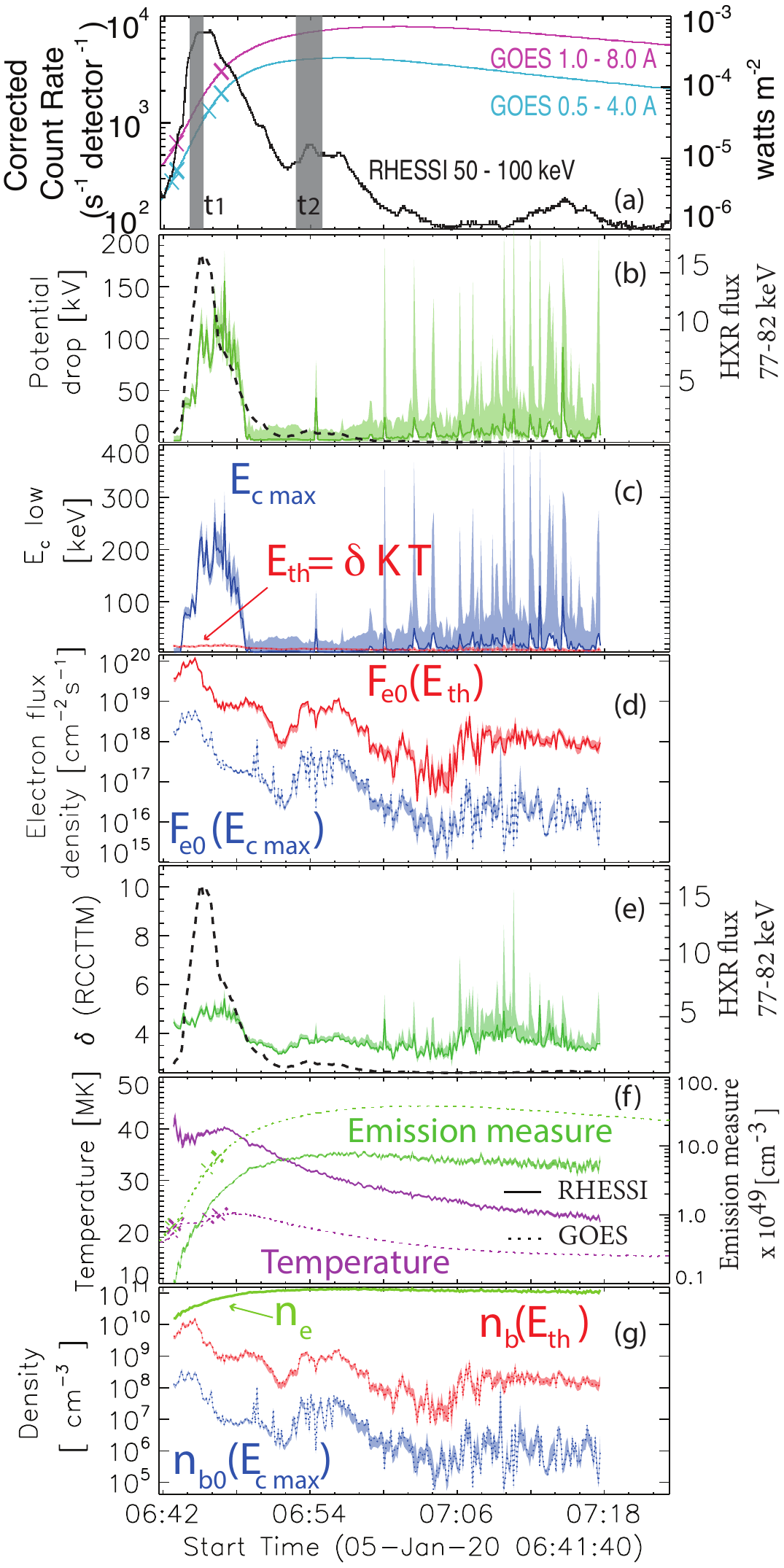}
\end{minipage}\hfill
 \begin{minipage}[c]{0.32\textwidth}
\caption{Time evolution of fit parameters from 2005-Jan-20. Same as figure~\ref{fig:timeevo}, \href{https://hesperia.gsfc.nasa.gov/collaborate/malaouia/public_html/movie4html/source20050120.pdf}{source sizes} can be found online. }
\label{fig:timeevo2}
 \end{minipage}
 
\end{figure*}

Figures ~\ref{fig:timeevo} and ~\ref{fig:timeevo2} show the time evolution of the deduced 2005-Jan-19 and 2005-Jan-20 flare parameters. We fit RHESSI spatially-integrated spectra over 8 s time bins between 08:12 and 08:29 UT for the \href{https://hesperia.gsfc.nasa.gov/collaborate/malaouia/public_html/movies4html/20050119/movie.html}{2005-Jan-19} flare and between 06:42 and 07:18 UT for the  \href{https://hesperia.gsfc.nasa.gov/collaborate/malaouia/public_html/movies4html/20050120/movie.html}{2005-Jan-20} flare. We show results using detector 4, but we have fitted the data using detectors 1 and 5 separately to check for consistency. We corrected for pulse pile up \citep{2002SoPh..210...33S} and isotropic albedo using the default parameters in OSPEX.  \\

These two flares have some of the highest fitted potential drops in our sample, and both exhibit potential drops higher than 100 kV. These high potential drops correspond to the last HXR bursts between 08:24 and 08:28 UT in 2005-Jan-19 flare, and the first HXR burst between 06:42 and 06:50 UT in 2005-Jan-20 flare.\\

Panel (a) in figures  ~\ref{fig:timeevo} and ~\ref{fig:timeevo2} shows the nonthermal RHESSI lightcurves integrated over 50 to 100 keV in black and the thermal GOES lighcurves at 1.0 - 8.0 $\AA$ and 0.5 - 4.0 $\AA$ in purple and light blue, respectively. The grey shaded areas labelled t1 through t5 are time intervals where the source sizes have been calculated. These source images can be found online by clicking on the flare date in Table~\ref{tab:sourcestable}. The x symbol in the GOES lightcurves represents the beginning or end of bad data.\\

Panel (b) shows the time evolution of the potential drop from the loop top to the thick target, V$_{TT}$. When the potential drop is negligible, such as during 2005-Jan-19 flare before 08:17 UT, the return current collisional thick target model fit becomes the standard collisional thick target model fit. The burstiness after 06:54 UT during 2005-Jan-20 is unlikely to be real because the spectra could be fitted with a single power-law with the sharp low-energy cutoff close to and below the transition energy between the isothermal and nonthermal components: in other words, the nonthermal x-ray spectrum does not flatten at lower energies.\\

Panel (c) shows the time evolution of the upper and lower limits of the cutoff energy, as explained in section ~\ref{constraints}. Note that everywhere in the paper, results using the lower limit of the cutoff energy are represented in red/pink and results using the upper limit of the cutoff energy are represented in blue/light blue.  \\

Panel (d) shows the upper and lower limit of the electron flux density in electrons cm$^{-2}$ s$^{-1}$ at the acceleration site. Since the spectral fits provide the differential electron flux (electrons s$^{-1}$ keV$^{-1}$) at 50 keV, the total electron flux density at injection is given by equation (~\ref{(2)}), where E = 50 keV, divided by the area of the footpoints A$_{FP} $, as calculated in section ~\ref{source sizes}. This gives
\begin{equation}{F_{e0}(E_{c\, max}) ={ {F(50, x_{TT}) \over {A_{FP} {(\delta -1)}}} {{(50+ e\,V_{TT})}^{\delta} \over {(E_{c\, max})}^{(\delta-1)}}} }  
 \label{eq:fe0}\end{equation} 

 where A$_{FP}$ is the total footpoints area, x$_{TT}$ is the distance from the loop top to the thick target, and E$_{c\, max}$ is the injected low-energy cutoff. Similarly, in the lower limit of the cutoff energy E$_{th}$= $\delta$ k T, the total electron flux density injected at the looptop is 
\begin{equation}F_{e0}(E_{c \,min} = E_{th})= { {F(50, x_{TT}) \over {A_{FP} {(\delta -1)}}} {{  {50^{\delta}} \over {E_{th}^{(\delta-1 )}}} }} 
 \label{eq:fe0eth}\end{equation}.

  \twocolumn

\newpage
\onecolumn

  \twocolumn

\onecolumn

Panel (e) shows the electron spectral index in green and the HXR flux in units of photons cm$^{-2}$ s$^{-1}$ keV$^{-1}$ in the energy bin 152-162.5 keV for 2005-Jan-19 and in the energy bin 77-82 keV for 2005-Jan-20. \\

\begin{figure}
\centering
\includegraphics[width=6cm]{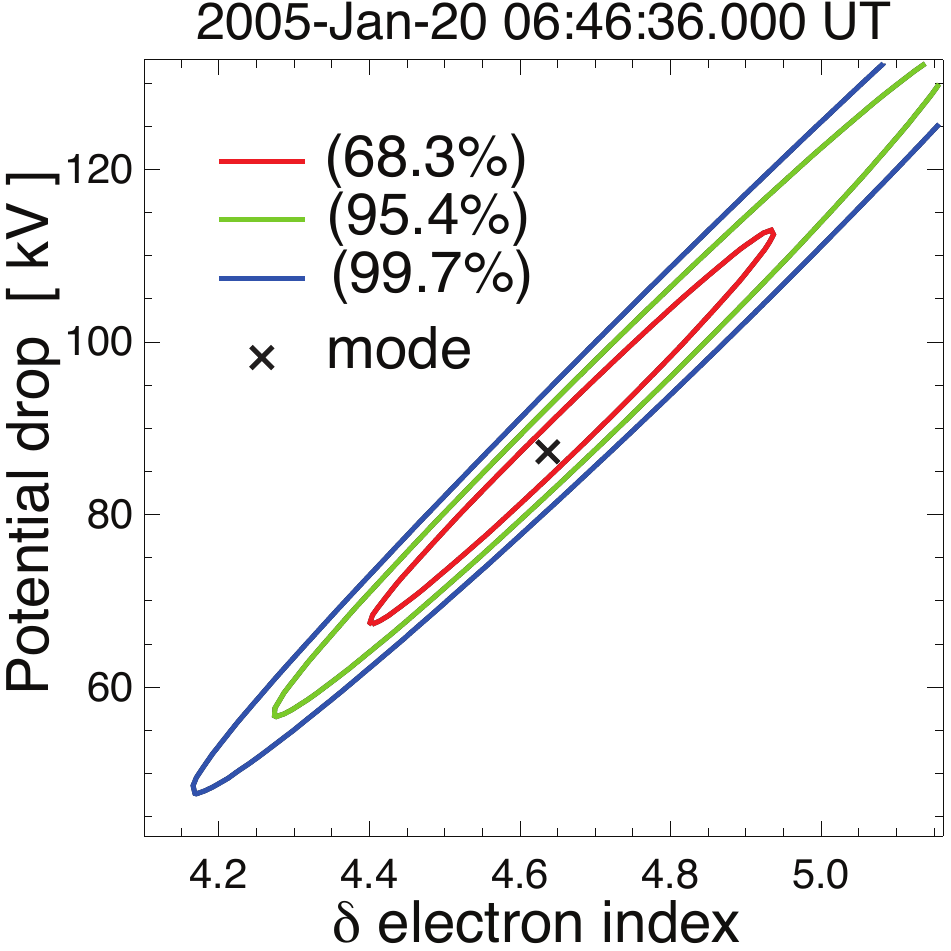}
\caption{ Potential drop vs electron spectral index for the 8-s interval starting at 2005-Jan-20 06:46:36 UT. The contours represent the confidence intervals. }
\label{fig:chisqmap}
\end{figure}

The spectral index and the potential drop are highly correlated, and increasing one parameter also increases the other to obtain the best fit. A higher spectral index would require a higher potential drop to explain the same flattening in the photon spectrum. Figure ~\ref{fig:chisqmap} shows the dependence of the potential drop on the spectral index at 2005-Jan-20 06:46:36 UT. The contours are the 68.3$\%$ (red), 95.4$\%$ (green), 99.7$\%$ (blue) confidence intervals. \\

There is an anti-correlation between the spectral index and the HXR flux during 2005-Jan-19  before 08:23 UT, where the potential drop is negligible. The rank correlation coefficient is $\rho$ = -0.80. This is known as soft-hard-soft (SHS) behavior  \citep[e.g.,][]{1969ApJ...155L.117P, 1977ApJ...211..270B, 1987ApJ...312..462L, 2004A&A...426.1093G}, and is thought to be a feature of the acceleration mechanism.  After 08:24 UT, there is no correlation or anticorrelation between the spectral index and HXR flux. However, there is a correlation between the potential drop and the HXR flux with the rank correlation coefficient equal to 0.59. This coefficient increases to 0.78 if the spectral index is fixed at 4 and the potential drop is allowed to change. Note that fixing the potential drop at 0 does not provide acceptable fits, as there is a flattening at lower energies in the photon spectra after 08:24  UT. However, fixing the potential drop at 50 kV, which is an average value of the fitted potential drop when $\delta$ is fixed at 4, and fitting the spectral index provides comparable or worst fits as compared to the case where the spectral index is fixed at 4. The Spearman correlation coefficient between the HXR flux and the spectral index is $\rho$= -0.68. Therefore, the higher flattening at the HXR peak time, i.e., SHS behavior is consistent with the return-current potential drop, a transport effect, as opposed to the SHS behavior associated with the acceleration process before 08:23 UT. During 2005-Jan-20, there is no correlation or anticorrelation between the spectral index and the HXR flux, but there is a correlation ($\rho$ = 0.87) between the potential drop and the HXR flux during the strongest HXR peak between 06:42 and 06:49 UT. This is consistent with SHS behavior at the thick target only. \cite{2005A&A...439..737B} investigated whether SHS behavior is a feature of the acceleration region or a transport effect by performing imaging spectroscopy of both the coronal and footpoint sources. They found that in most cases there is a SHS behavior in the coronal source and concluded that the SHS behavior is a property of the acceleration region. They also found one event, the 01-Nov-2003, in which the SHS behavior is a feature of both the coronal source and the footpoints. \\

Panel (f) is the time evolution of the best-fit temperature in purple and the emission measure in green. The solid curves with uncertainties represent RHESSI parameters, and the dashed curves are GOES parameters. The temperature peaks before the emission measure and decays faster in both RHESSI and GOES data, and in both flares.\\

Panel (g) is the time evolution of the beam density at the loop top and the thermal background density. The background density is deduced from the thermal fit and the recontructed thermal volume $\Omega$ as calculated in section ~\ref{source sizes}, \begin{equation}n_{e} = {\sqrt{{EM} \over {\Omega}}}\, \, ,\label{eq:density_bkg}\end{equation} where EM is the emission measure in cm$^{-3}$. The number density in the beam is calculated from the nonthermal fit parameters: \begin{equation}n_{b0}=  \int f(v) \,d^3 v\,=\int_{E_{c0}}^\infty{F(E)\over v }dE \label{eq:beam_density}\end{equation} where $v$ is the speed of an electron. Using equation (~\ref{eq1}), this gives \begin{equation}n_{b0}={ \sqrt{m_e \over 2} {{(\delta -1)} \over {(\delta -{1 \over 2})}}E_{c0}^{-1/2} F_{e0}, }\label{eq:density_beam}\end{equation} where E$_{c0}$ is the cutoff energy at the looptop and $m_e$ the electron mass. The beam density at the footpoints is equal to that at the looptop in the upper limit of the low-energy cutoff because no electrons are thermalized. In the lower limit of the low-energy cutoff, since electrons are thermalized, the beam density at the footpoints is always lower than that at the looptop. Note that the electron beam density is usually lower than the background density by a few orders of magnitude in both limits of the low-energy cutoff in both flares, except for the times between 06:42 and 06:45 UT during 2005-Jan-20 in the lower limit of the low-energy cutoff, when the two densities are on the same order of magnitude. When the density in the background is significantly higher than the density in the nonthermal beam, there are enough thermal electrons to form the return current. Section~\ref{energy_total} further constrains the cases where return-current losses is the preferred model to explain the photon spectral flattening. \\


\section{Statistical results from all flares}
\label{stat}
\begin{figure}[bh!]
\centering
\includegraphics[width=16.5cm]{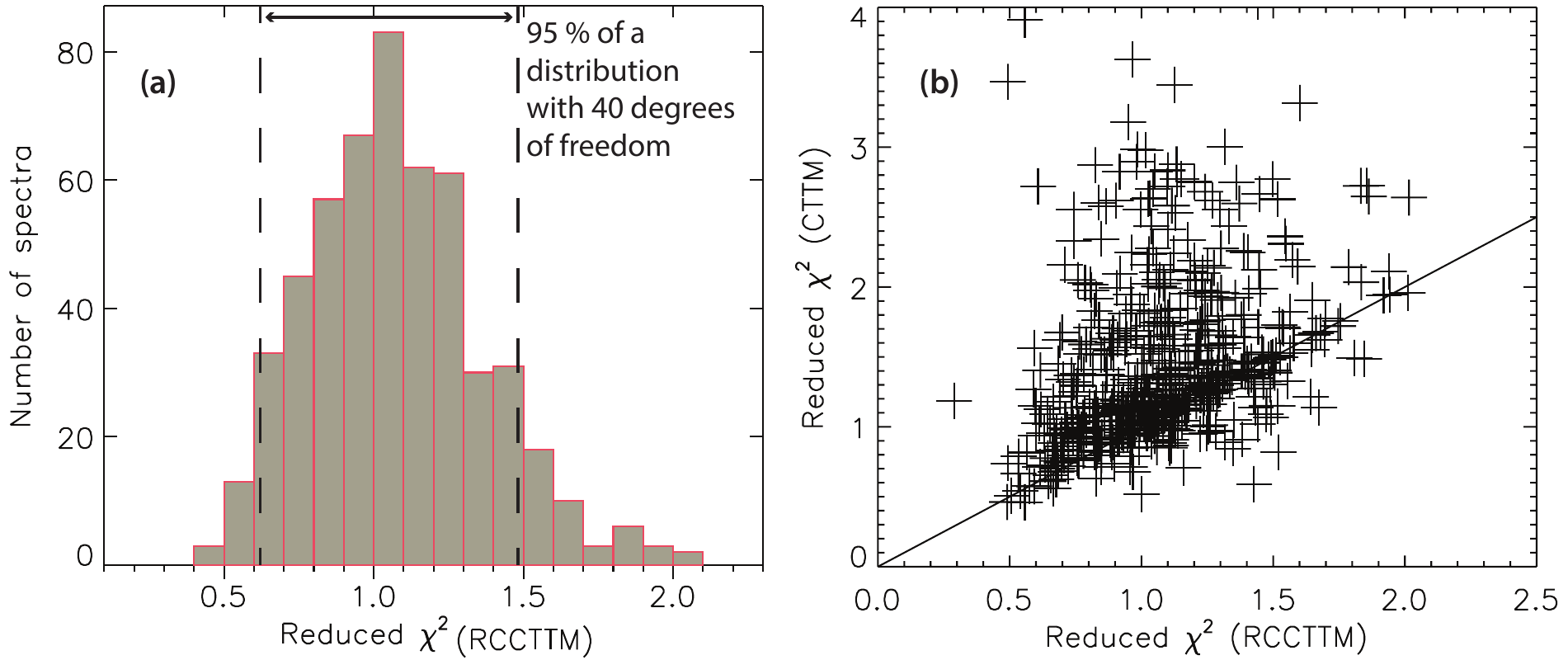}
\caption{Reduced $\chi^2$ statistics for 528 spectra from 18 flares, including all spectral fits with the potential drop greater than 10 kV and 2 $<\delta<$ 7. This reduces the sample from 1154 spectra to 528 spectra from the 18 flares. \textit{(a):} Histogram of reduced $\chi^2$ values from the RCCTTM fits.\textit{(b):} Scatter plot of reduced $\chi^2$ values from CTTM fits vs RCCTTM. The solid line is where the reduced $\chi^2$ from CTTM and RCCTTM are equal.  }
\label{fig:chisq}
\end{figure}

In the previous section, we showed the time evolution results from two flares as examples and to investigate time-dependent patterns such as SHS originating from behavior in the acceleration region (anti-correlation between the electron spectral index and the HXR flux) or at the footpoints caused by the time evolution of the potential drop (correlation between the HXR flux and potential drop). We are now interested in statistical results from all spectra with a strong enough flattening, which is defined as spectra with potential drops higher than 10 kV and an electron spectral index 2$<\delta<$7.\\

Panel (a) in figure~\ref{fig:chisq} shows a histogram of reduced $\chi^2$ values using the RCCTTM. The two vertical dashed lines show the 2$\sigma$ or 95$\%$ confidence interval for 40 degrees of freedom (DoF). The number of DoF depends on the number of energy bins, and since the upper limit on the photon energy to fit depends on the background level, the fitted spectra have different numbers of DoF up to 44 and down to 20. \\
\begin{figure}
\centering
\includegraphics[width=16cm]{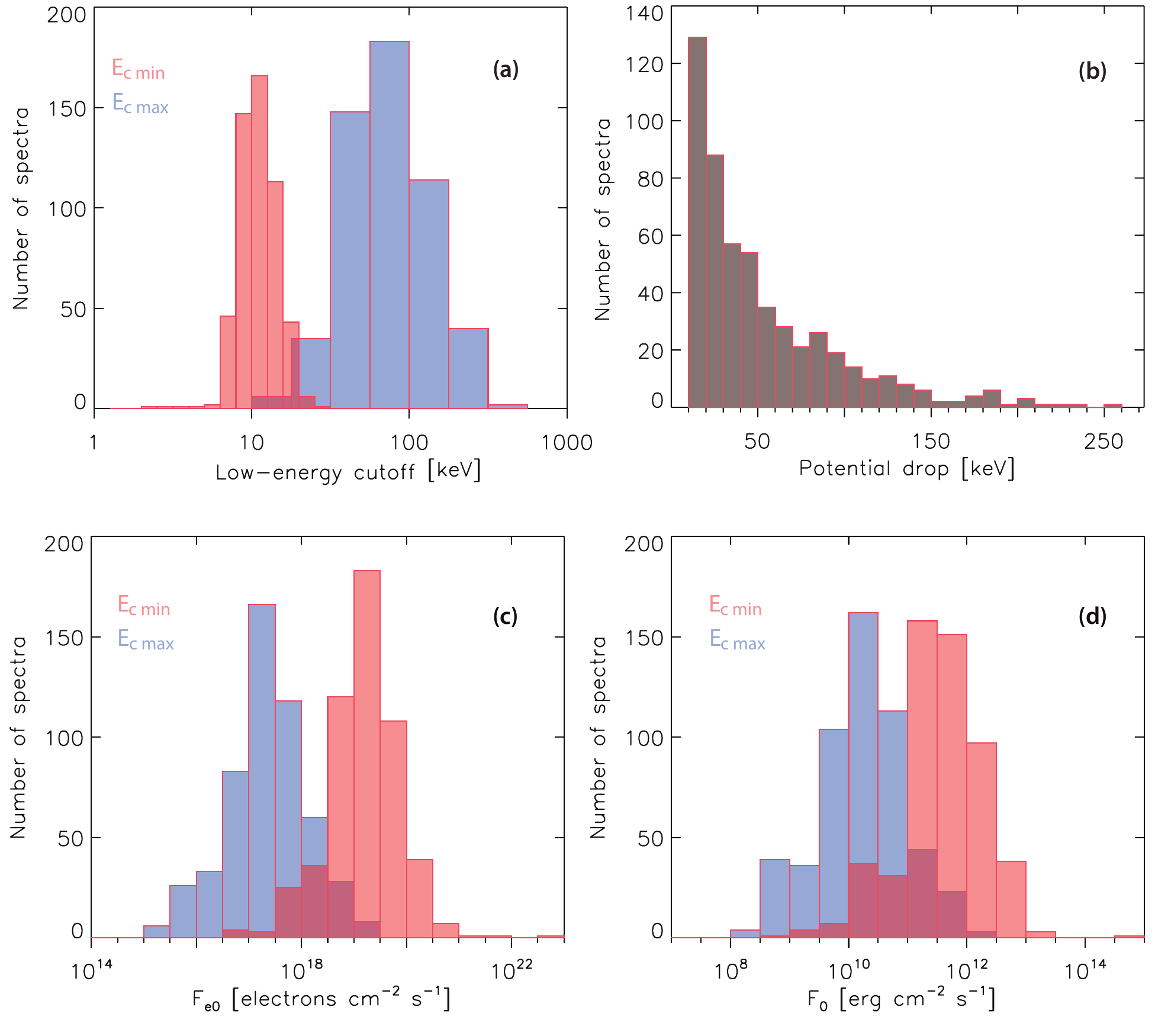}
\caption{Frequency distributions using only spectra with potential drop greater than 10 kV and 2 $<\delta<$ 7. \textit{(a):} Histogram in log scale of the upper and lower limits of the cutoff energy at the acceleration region in blue and red, respectively. \textit{(b):} Histogram of the fitted potential drop. \textit{(c):} Histogram of the injected electron flux density at the acceleration region in the upper/lower limits of the low-energy cutoff in blue/red in units of electrons cm$^{-2}$ s$^{-1}$.\textit{(d):} Histogram of the injected energy flux density at the acceleration region in the upper/lower limits of the low-energy cutoff in blue/red in units of erg cm$^{-2}$ s$^{-1}$ }
\label{fig:histo}
\end{figure}

Panel (b) is a scatter plot of $\chi^2$ values from the standard collisional thick target model versus $\chi^2$ from the return current model. In the CTTM we use the best $\chi^2$ value from either a single power-law with a sharp low-energy cutoff or a double power-law with a sharp low-energy cutoff. The solid line is where $\chi^2$(CTTM)= $\chi^2$(RCCTTM). In most cases, the two values are scattered around the solid line, meaning that spectra could be fitted equally well with both models, but there are many cases where the $\chi^2$(CTTM)$>$2, and the RCCTTM fits the spectra better. Remember the flares were chosen because they exhibit a strong flattening that could be well-fitted with a single power-law and a high value of the low-energy cutoff, or double power-law and a sharp low-energy cutoff at the HXR peak time. From our selection criteria, the scatter plot shows that spectra from these flares are usually well-fitted with the RCCTTM over the entire impulsive phase. However, the CTTM fits are not acceptable over the entire impulsive phase. The only exception we found is the   \href{https://hesperia.gsfc.nasa.gov/collaborate/malaouia/public_html/movie4html/20020822/movie.html}{2002-Aug-22} flare for which the return current fits were significantly worse than a broken power-law fit. We think the  \href{https://hesperia.gsfc.nasa.gov/collaborate/malaouia/public_html/movie4html/source20020822.pdf}{complexity of the source}, which shows at  least four footpoints, is likely responsible; and the double power-law fit could be due to different populations of electrons accelerated on different loops.\\

Figure ~\ref{fig:histo} shows occurrence frequencies of return-current model parameters, again using spectra with a potential drop greater than 10 kV and 2$<\delta<$7, reducing the sample of spectra from 1154 to 528 from 18 flares. \\

Panel (a) shows the upper (blue) and lower (red) limits of the low-energy cutoff at the injection site near the looptop. Note that, as in figures ~\ref{fig:RCCTTM_CTTM_20050120} and ~\ref{fig:RCCTTM_CTTM_20050119}, the values of the upper limit of the cutoff energy are higher than obtained with the standard collisional thick-target model. This is explained by the assumption of the models. The electron distribution just above the thick target is assumed to be unchanged from the acceleration region in the CTTM, because electrons do not lose any significant amount of energy until they reach the thick target, whereas in the RCCTTM the observed spectrum at the footpoints is assumed to have evolved from the injected single power-law due to the potential drop from the acceleration region to the footpoints. The injected E$_c$ is then higher than the observed thick target E$_c$. \\

Panel (b) shows the potential drop histogram, where the highest potential drop is about 260 kV. This upper limit is determined by the maximum fitted photon energy of 300 keV, because we require that part of the power-law as well as the flattening at lower energies to be fitted. The lower limit is 10 kV by definition. The peak of the distribution is between 10 and 20 kV because there are more spectra with lower flattenings. The higher flattenings correspond to the highest potential drop values, which correspond to the HXR peak times. There is a weak correlation between the HXR flux and the potential drop: for example the rank correlation coefficient between the photon flux in the energy bin 108-116 keV and the potential drop is $\rho$ = 0.51. This indicates that in our sample of flares, which were chosen because they have strong breaks at the HXR peak time, exhibit stronger breaks at peak time as compared to valleys in the lightcurve. \\

Panels (c) and (d) show the electron number and energy flux density histograms in both limits of the low-energy cutoff. The most probable value of the energy flux density is 3$\times$10$^{11}$ erg cm$^2$ s$^{-1}$ in the lower limit of E$_c$ and 1$\times$10$^{10}$ erg cm$^2$ s$^{-1}$ in the upper limit of E$_c$. This result is biased by the fact that larger, more powerful flares, which also exhibit higher values of the potential drop, have longer impulsive phases and hence more spectra in our sample. \\
\section{Coronal and beam/return-current parameters}
\label{section5}
  \subsection{Return current electric field}

 \cite{2012ApJ...745...52H} derived analytical results in the simple steady-state 1D case where electrons lose energy in the corona through deceleration by the co-spatial return-current electric field. All electrons lose the same amount of energy, determined by the potential drop which is proportional to the electric field magnitude: ${{dV} \over {dx}} = e \mathscr{E}_{rc} (x)$, where V is the potential drop, x the distance from the acceleration region (loop top or cusp), e is the charge of the electron, and $\mathscr{E}_{rc}$(x) is the return current electric field magnitude at position x.\\
 
The return current electric field is proportional to the current density through Ohm's law: \begin{equation}
{{\mathscr{E}}}_{rc}(x) = \eta(x) {J}_{rc}(x)= \eta(x) {J}_{direct}(x)= \eta e F_{e}(x)\label{ohm}\end{equation} where $\eta$ is the resistivity of the plasma and J$_{direct}$ = e F$_{e}$(x), where F$_e$ is the flux density of the beam of accelerated electrons (electrons cm$^{-2}$ s$^{-1}$).\\
 
The electric field strength is constant along the loop as long as the electron flux density and the resistivity are constant. However, if the energy of an electron is reduced to a few times the thermal energy, assumed to be equal to E$_{th}$= $\delta$ k T (fitted), as estimated by \cite{2015ApJ...809...35K}, the electron is lost from the nonthermal beam to the thermal background. This is highly sensitive to the low-energy cutoff because injected electrons with an energy equal to the low-energy cutoff are the first to be lost from the nonthermal beam. This happens when E$_c -$V(x) $\le$ E$_{th}$.\\

Two main regimes arise where the beam is decelerated by the return current:\\

\noindent (1) All electrons have energies above the thermal background energy E$_{th}$ until they reach the thick target, where the density is high enough to stop them through Coulomb collisions. In this case, with constant resistivity along the loop, the electric field remains constant and decelerates the beam at the same rate along the loop.\\

\noindent (2) Some electrons are lost from the beam (thermalized) and the electric field strength decreases downward. The magnitude and impact of the return-current electric field is then greater higher in the loop.\\

\noindent Let's explore what happens in the upper and lower limits of the cutoff energy. \\
At $E_{c \, \,max}= E_{c\, \, TT} +e\, V_{TT} > E_{th} +e\,V_{TT}$, no electrons are lost from the injected beam and the electric field is constant along the loop and given by:
\begin{equation} \mathscr{E}_{rc}(E_{c\, max})=\mathscr{E}_{rc0}(E_{c\, max})= {{V_{TT}} \over { x_{TT}}}.\label{eq:erc2}\end{equation}
\\

At E$_{c \, min}$ = E$_{th}$= $\delta$ k T, electrons are lost from the beam as soon as they are injected. Therefore, the electric field at the loop top, assuming the temperature is constant along the loop, is given by the potential drop corrected for the electron losses from the injected beam. We start with equation (13) in  \cite{2012ApJ...745...52H} \begin{equation} {dV \over dx}= \mathscr{E}_{rc}(x)=e \, \eta\, F_e(x)\,  , \label{eq:erc}\end{equation} and recognizing that the electron flux density is given by $F_e(x)= F_{e0} \, {(1\, + {e V(x) \over {\delta kT}})}^{(1-\delta)}$, we write equation~(\ref{eq:erc}) as \begin{equation}{dV \over dx}=e\, \eta\, F_{e0} \, {(1+{e\, V \over {\delta kT}} )^{(1-\delta)}} \label{eq:10}\end{equation}

which is a separable differential equation. After solving and integrating equation~(\ref{eq:10}) over the length of the loop, we obtain the return-current electric field in terms of observable quantities:

      \begin{equation} \mathscr{E}_{rc0}(E_{c\, min})= {( (1 + {{e\,V_{TT}}  \over  {\delta k T}} )^{\delta} -1)\times  { {kT}  \over { x_{TT}}}} \end{equation} 
This is the maximum return-current electric field strength at the looptop, which decreases down the length of the loop.  \\

At the thick target, the return-current electric field strength decreases by the same amount the electron flux density decreases, given by the factor ${(1+ {e\, V_{TT} \over {\delta kT}})}^{(1-\delta)}$.  \begin{equation} \mathscr{E}_{rc}(E_{c\, min},x_{TT})= {( (1 + {{e\, V_{TT}}  \over  {\delta k T}} )^{\delta} -1)  {(1+{e\, V_{TT} \over {\delta kT}})}^{(1-\delta)}  \times  { {kT}  \over { x_{TT}}}} \label{eq:erc1} \end{equation}

\begin{figure}
\centering
\includegraphics[width=7.5cm]{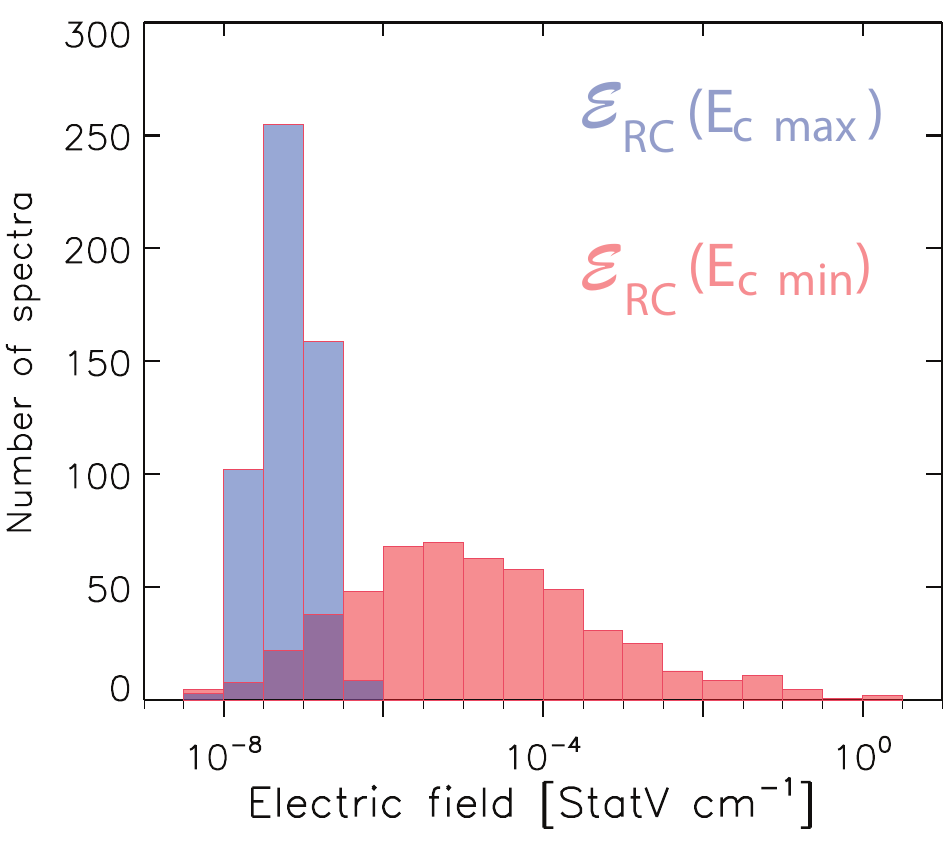}
\caption{Histogram of the return current electric field in the upper (lower) limit of the cutoff energy in red (blue).} 
\label{fig:efield_rc}
\end{figure}

\noindent Figure ~\ref{fig:efield_rc} shows a histogram of the return-current electric field at the looptop in both limits of the low-energy cutoff.  Only results from spectra with 2$<\delta<$7 and V$_{TT}\, >$10 kV are plotted. The mean value of the electric field is $\mathscr{E}_{rc0}(E_{c\, \, max})=$3.2$\times$10$^{-8}$ StatV cm$^{-1}$ and $\mathscr{E}_{rc0}(E_{c\, \, min})=$3.2$\times$10$^{-6}$ StatV cm$^{-1}$. \\

\subsection{Is the resistivity in the corona consistent with classical Spitzer values?}
\label{resistivity section}
The resistivity in the corona is deduced from Ohm's law: equation (~\ref{ohm}). 
 The electron flux density at the thick-target F$_{e}$(x=x$_{TT}$) is given by the integral of the electron distribution at position x$_{TT}$: $F_{e}(x=x_{TT})= \int_{E_{c \,\,TT}}{F(E,x=x_{TT})} \, dE$ with the electron distribution given by equation~\ref{(2)}. The electron flux density becomes 
For  $E_c = E_{c\, max}$, 

 $F_{e}(x_{TT})=  F_{e0} $ and for $E_c= E_{th}=\delta k T$, $F_e(x_{TT})\,= F_{e0}{(1+{{e\, V_{TT}} \over {E_{th}}})^{1-\delta}}$
 \\

And the resistivity is, from equations ~\ref{ohm}, ~\ref{eq:erc2} and ~\ref{eq:erc1}:
\begin{equation}
\eta_{corona} =
 {{V_{TT}} \over {e  F_{e0} x_{TT}}} \end{equation}

 \begin{equation}
\eta_{corona} ={{k T} \over {e F_{e0} x_{TT}}}{[{(1+{ {e\, V_{TT}} \over {\delta k T}})}^{\delta}-1 ]} 
\label{eq_resistivity}
 \end{equation}\\

\twocolumn

\begin{figure*}[t]
  \begin{minipage}[c]{.5\linewidth}
    \null
\includegraphics[width=\textwidth]{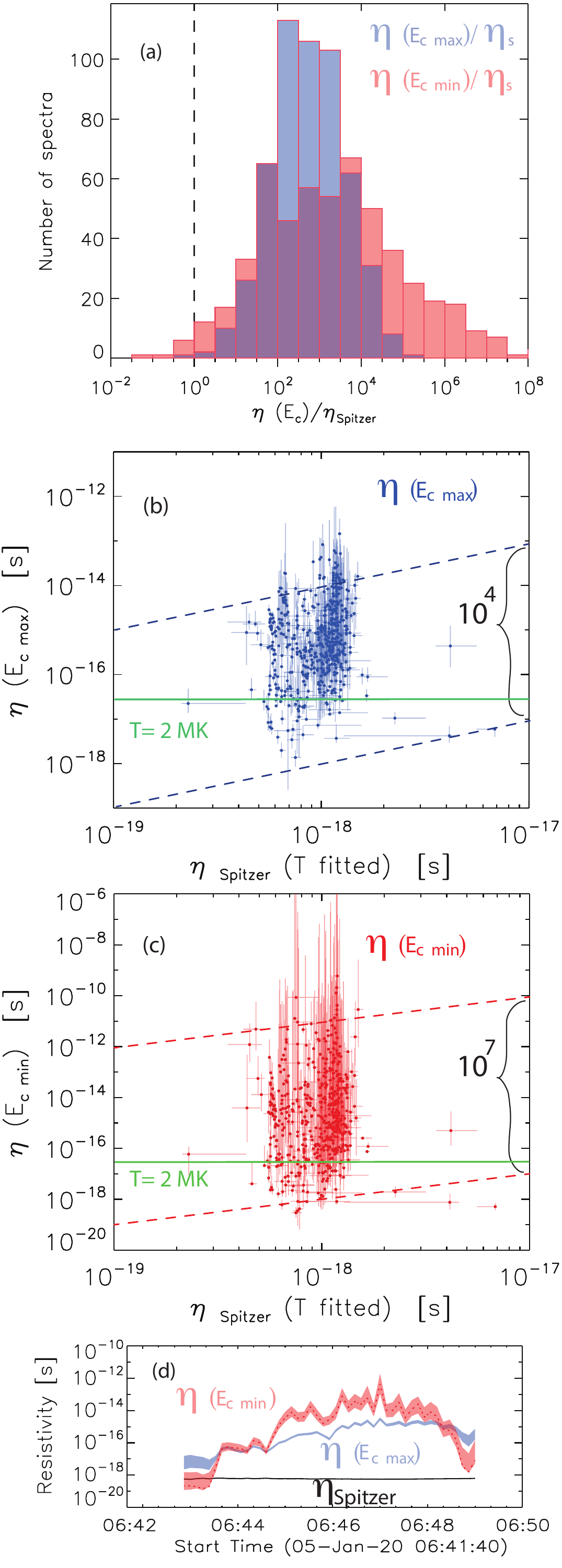}
\end{minipage}\hfill
 \begin{minipage}[c]{0.32\textwidth}
\caption{ Derived resistivity values from Ohm's law. (a): Histogram of the ratio of derived resitivity to the Spitzer resistivity in the upper (lower) limit of the cutoff energy in blue (red) for 528 spectra. The dashed vertical line is where this ratio is equal to 1.   Note that all but one data point have higher resistivities than the Spitzer resistivity in the upper limit of E$_c$, with a mean value at 10$^2$, i.e., the derived resistivity is typically two orders of magnitude higher than Spitzer values.  (b) and (c): Scatter plots of the derived resistivity in the upper (b) and lower (c) limits of the low-energy cutoff vs the classical Spitzer resistivity. The error bars are 67$\%$ confidence intervals. (d) Time evolution of resistivity in the upper (lower) limits of the cutoff energy in blue (red) for the first peak of 2005-Jan-20; the Spitzer resistivity at the fitted temperature is plotted in black. }
\label{fig:resistivity}
 \end{minipage}
 
\end{figure*}
\onecolumn

We compare this derived resistivity to the classical Spitzer resistivity \citep[][]{1962pfig.book.....S}, which is given by: \begin{equation} \eta_{S}= {{m_e \nu_{ei}} \over {1.96 n e^2   } } = 1.46 \times 10^{-7} {\lambda \over 20} T^{-3/2} [s]\label{eq:spitzer}\end{equation} where  $\nu_{ei}$ is the thermal electron-ion collision frequency, and T is the temperature in K and $\lambda$ is the Coulomb logarithm given by: \\
  for   $T < 1.16\times 10^5 \, Z^2$ 
\begin{equation}
\lambda =
 {8.96 - ln{(Z\, n^{1/2} \, T^{-3/2})}} \end{equation}
 and  for   $T > 1.16\times 10^5 \, Z^2$ 
   \begin{equation}  
\lambda={14.6+ln{(n^{-1/2} T)}}  
\label{eq:lambda}
\end{equation}

Figure ~\ref{fig:resistivity} (a) shows the ratio of the derived resitivity to the classical Spitzer resistivity at the fitted temperature for the 528 spectra. The dashed vertical line is where the ratio is equal to 1. In the upper limit of E$_c$, the derived resistivity is higher than the Spitzer values by up to 5 orders of magnitude, for all but one case which is equal to the Spitzer value within the 67$\%$ confidence interval. The most probable value is 2 orders of magnitude higher than Spitzer values.  In the lower limit of E$_c$, the resistivity in the corona could be as high as 8 orders of magnitude higher than Spitzer values at the fitted temperature, with the most probable values between 2 and 4 orders of magnitude higher than Spitzer values. In the lower limit of the low-energy cutoff, the resistivity derived from Ohm's law is lower than the Spitzer resistivity at the fitted temeprature in 9 spectra from 3 flares. This is physically implausible and could be explained by either the presence of a higher temperature component than the isothermal temperature fit, or the low-energy cutoff $\delta$ k T is simply too low.\\

Figures ~\ref{fig:resistivity} (b) and (c) show the derived resistivities plotted againt the Spitzer values for E$_{c\, max}$ (c) and E$_{c\, min}$ (d). The narrow range of values of the Spitzer resistivity is due to the narrow range of highest temperatures fitted with RHESSI, which is between 17 MK and 46 MK for the flares in our sample. \\

Figure ~\ref{fig:resistivity} (d) shows the time evolution of the resistivity during the HXR peak between 06:43 UT to 06:49 UT in 2005-Jan-20. The resistivity for E$_{c\, max}$ is increasing from one order of magnitude above the Spitzer value to more than 3 orders of magnitude above Spitzer resistivity, because the potential drop is increasing and the electron flux density decreases by more than an order of magnitude. The temperature is between 37 and 42 MK, so the Spitzer resistivity is nearly constant, and the enhanced resistivity is independent of the temperature. There is no clear correlation between the HXR flux and the resistivity. \\

Although all derived resistivities are higher than the Spitzer value in the upper limit of E$_c$, some of these resistivities could be classical. For example, the nonthermal beam could be streaming down a nearby loop with a lower temperature than the fitted temperature with RHESSI, or another possibility, although less likely, most of the loop length is dominated by a lower temperature component. In these cases, the classical Spitzer resistivity is higher because it scales as T$^{-3/2}$. As an example, the corresponding temperature of a resistivity $\eta= 4.8 \times \,10^{-17}$ s, about where the green line in figure~\ref{fig:resistivity} (b) and (c) is, assuming it is classical and equal to the Spitzer value in a fully ionized atmosphere, and with density 10$^{10}$ cm$^{-3}$, is 2 MK. This is a plausible temperature in the corona and could be the temperature in a nearby cooler loop, where electrons are newly accelerated. However, most of the derived resistivity values are higher, which means that most values of the derived resistivities are anomalous.\\

In the lower limit of E$_c$, the assumption of classical resistivity becomes more difficult to defend. The derived resistivity is up to eight orders of magnitude higher than Spitzer values, which means that the temperature should be twelve orders of magnitude lower than the fitted RHESSI temperatures, assuming a fully ionized atmosphere: $\eta\propto T^{-3/2}$ which is equivalent to T$\sim 10^{-8\times {3 \over 2}}$ T$_{fitted}$.  Therefore, in most cases, anomalous resistivity is the best explanation for the derived resistivity values, in both limits of the low-energy cutoff assuming Ohm's law, equation~\ref{ohm}, applies.\\

\section{Stability of the return-current}
\label{stability}
\subsection{Is the return current stable to the generation of current-driven instabilities?}
In this section we investigate the possibility that the beam-return-current system generates turbulence. Specifically, is the enhanced resistivity deduced from Ohm's law due to the return current itself? Several authors investigated current-driven instabilities in space plasmas. \cite{1977RvGSP..15..113P} examined the various mechanisms capable of producing anomalous resistivity, \cite{1985ApJ...293..584H} and references therein derived stability thresholds for the ion acoustic, electrostatic ion cyclotron, and Buneman instabilities. \cite{2002ASSL..279.....B} gives a review of current-driven instabilities.\\ 

\begin{figure}[bh!]
\centering
\includegraphics[width=18cm]{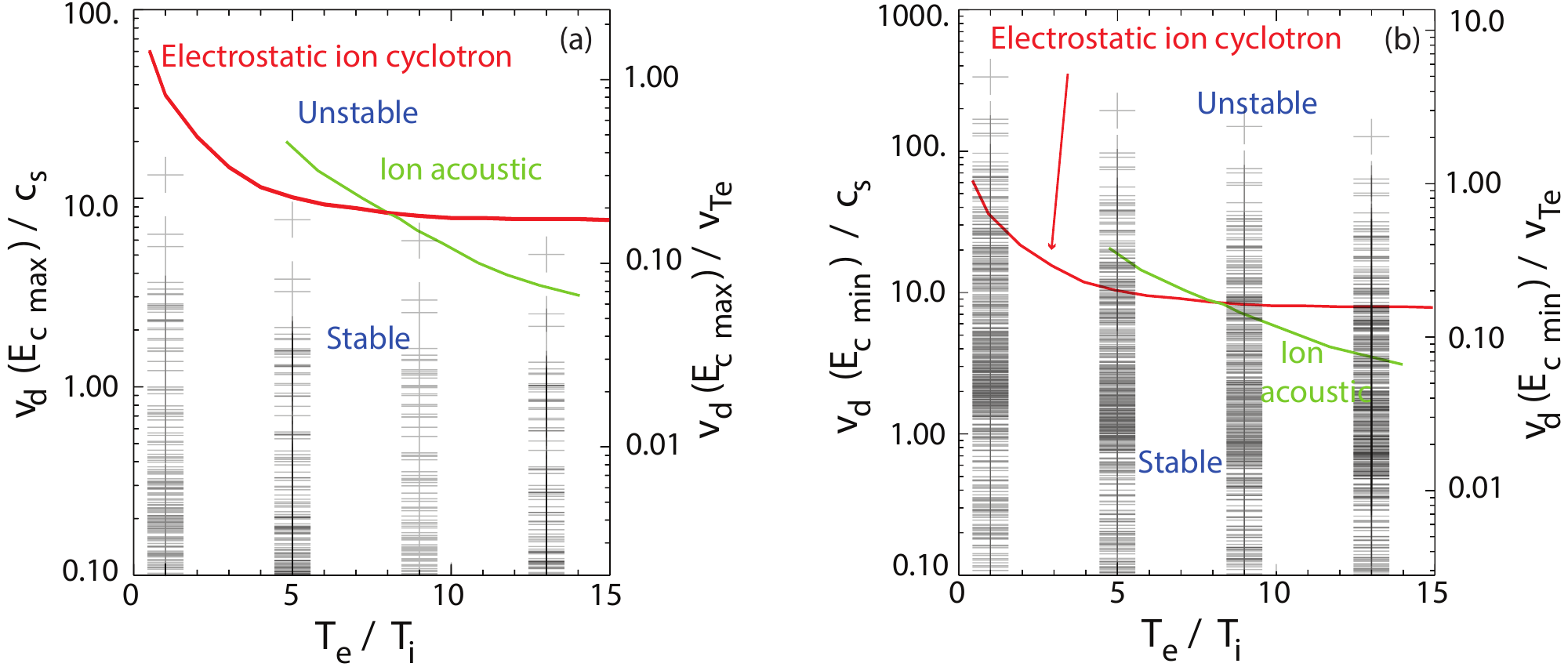}
\caption{Scatter plot of velocity ratios versus electron to ion temperature ratio. \textit{Left panel:}  Using the upper limit of the low-energy cutoff.\textit{Right panel:} Using the lower limit of the low-energy cutoff. The electron temperature is the fitted temperature from our sample.  \textit{Left axis:} Ratio of the drift velocity to the ion acoustic speed; \textit{Right axis:} Ratio of the drift velocity to the electron thermal speed. The red (green) curve is the limit to driving electrostatic ion cyclotron (ion acoustic) instability as derived in \cite{1985ApJ...293..584H}.  All 528 spectra with 2$\leq\delta\leq$7 and V$_{TT}\geq$10 are represented but for simplicity, left axes in both panels were cut off at $v_d \over c_s$ = 0.1. The lowest value of  $v_d (E_{c\, max}) \over c_s$ = 3$\times 10^{-4}$ and    $v_d (E_{c\, min}) \over c_s$ = 6$\times 10^{-3}$ }
\label{fig:holman}

\end{figure}

Figure~\ref{fig:holman}(a) shows the ratio of the return-current drift velocity, in the upper limit of the cutoff energy, to the ion sound speed assuming the electron temperature is 1, 5, 9, and 13 times the ion temperature. The red and green curves are taken from \cite{1985ApJ...293..584H} and represent the threshold of the electrostatic ion cyclotron (EIC) and ion acoustic (IA) instabilities, respectively. The return-current drift velocity is defined by \begin{equation} v_d (E_c)= {{J_{RC}(E_c)} \over {e n_{e}}}={{F_{e0}(E_c)} \over {n_{e}}}, \end{equation} and the ion acoustic speed is defined by $c_s={({{ k T_e} \over {m_i}} )  }^{1/2}$. The thermal velocity of the electrons is given by  $v_{Te}={({{ k T_e} \over {m_e} } )  }^{1/2}$. \\

 When $E_c =E_{c\,\, max}$, all data points fall in the stable regime if the electron and ion temperatures are equal and up to T$_e$ = 9 T$_i$, which indicates that the return current is stable to the generation of these instabilities. This implies that the anomalous resistivity is not generated by the return current itself, under the assumptions of the model.  As the ratio of the electron to ion temperature increases, more events are expected to be unstable, as shown by the one data point unstable to ion acoustic waves when this ratio is 13.\\

Similarly to the left panel, the right panel (figure~\ref{fig:holman} b) shows the ratio of the drift velocity in the lower limit of the cutoff energy to the ion sound speed assuming the electron temperature is 1, 5, 9, and 13 times the ion temperature. Some data points are unstable to the electrostatic ion cyclotron instability when T$_e$ = T$_i$, and as the ratio T$_e$/T$_i$ increases, more events become unstable to either EIC or IA instabilities. This is consistent with the results in sections~\ref{resistivity section} and ~\ref{dreicer section} , where the resistivity is more enhanced as compared to resistivity in the upper limit of E$_c$. Note that the cases where the ratio of drift velocity to the ion acoustic speed is higher than the EIC instability threshold do not correspond to the highest resistivities in figure~\ref{fig:resistivity} (d), the highest resistivity of this sample being 3.2$\times$10$^{-15}$s for a case where $v_d \over c_s$ =  94. Higher resistivity values correspond to return currents stable to EIC and IA instabilities. This is similar to the result for the upper limit of the low-energy cutoff, that the anomalous resistivity is not generated by the return current itself, under the 1D assumption of our model.\\ 

The Buneman instability is not of interest here because the threshold velocity is given by  $v_d\gtrsim 2 \,v_{Te}$. There are only up to five data points that reach this limit, and only in the lower limit of the low-energy cutoff, as shown on the right axis of panel b in figure~\ref{fig:holman}.\\

\cite{1977RvGSP..15..113P} showed that a stable beam cannot enhance the resistivity by more than a few percent. However, we have found that the return current is stable to current-driven instabilities and the resistivity is anomalous if Ohm's law is valid. In the next section, we explore the population of electrons responsible for the return current and review the assumption of deriving the resistivity from Ohm's law.\\

\subsection{What population of electrons carries the return current?} 
\label{dreicer section}
To derive the resistivity in the corona, Ohm's law was used, which has the inherent assumption that the bulk thermal electrons are responsible for carrying the return current. This explanation is only valid if the return current electric field strength is much lower than the Dreicer field. If the return current is carried by electrons that escape collisions, Ohm's law is invalid.\\ 

The classical Dreicer field is the critical electric field strength at which, for all of the electrons, acceleration by the imposed electric field is not balanced by the collisional drag. The classical Dreicer field is defined by the equation $e\mathscr{E}_D=m_e v_{Te} \nu_e$, with $\nu_e$ the electron thermal collision frequency, $v_{Te}$ the most probable velocity of thermal electrons, m$_e$ the mass of an electron, and $\mathscr{E}_D$ the classical Dreicer field \citep[][]{1959PhRv..115..238D}. In the presence of turbulence that enhances the resistivity compared to the Spitzer resistivity, the anomalous Dreicer field should be considered instead, and the electron thermal collision frequency is replaced by the anomalous collision frequency ${\nu_{an} = { \eta_{an} \over \eta_{S}} \nu_e}$, where $\eta_{an}$ is the anomalous resistivity.\\

\begin{figure}[bh!]
\centering
\includegraphics[width=17cm]{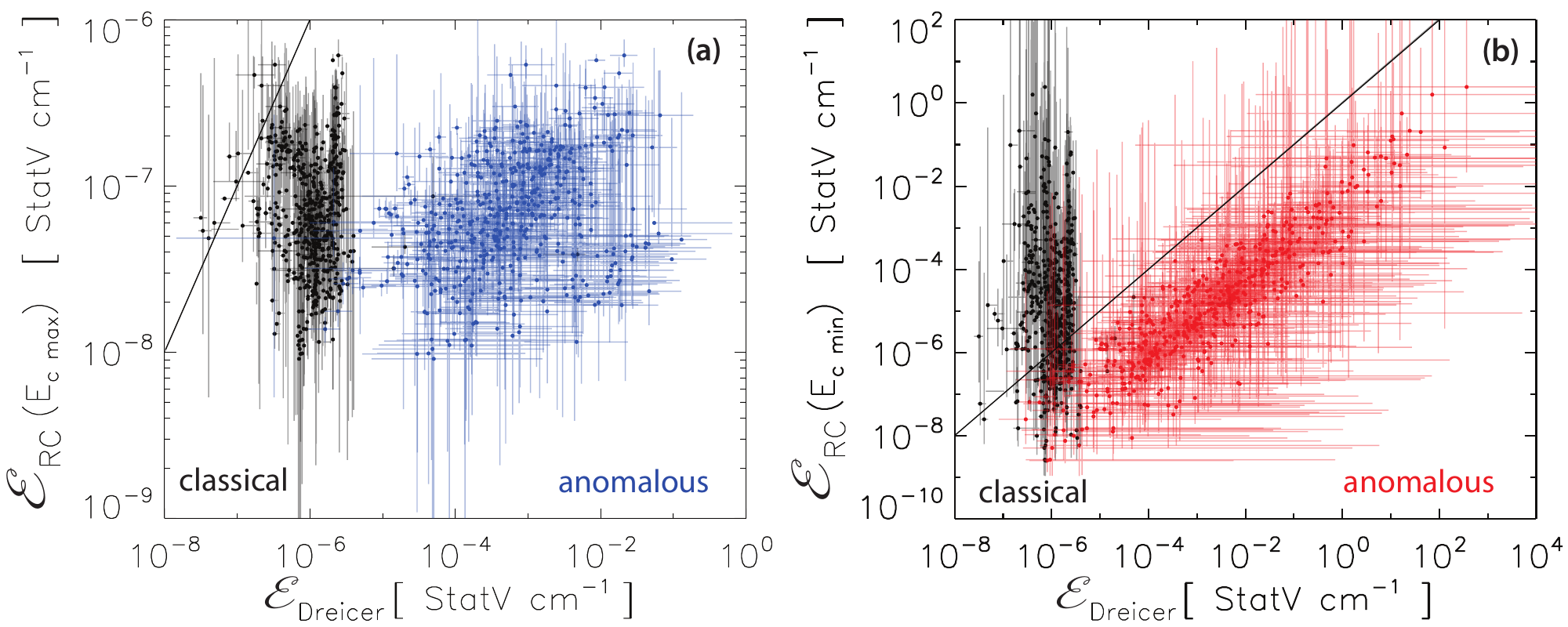}
\caption{Scatter plot of the return-current electric field strength versus the classical (black) and anomalous (blue or red) Dreicer field strength. The solid line is where $\mathscr{E}_{RC}=\mathscr{E}_{Dreicer}$ and the error bars represent the 67$\%$ confidence interval from 1000 Monte Carlo runs. (a) Using the upper limit of E$_c$. (b)Using the lower limit of E$_c$.}
\label{fig:dreicer}
\end{figure}

 In a strong electric field, equal to or stronger than the Dreicer field, all electrons are accelerated by the field and become \textit{runaway} electrons, and Ohm's law, which was used to calculate the resistivity, is not valid. If the return-current electric field is lower than the (classical or anomalous) Dreicer field, then only a fraction of the electrons, with a velocity higher than a critical velocity $v_{cr}$, will run away.\\

Figure~\ref{fig:dreicer} shows a scatter plot of the return-current electric field in the upper (lower) limit of E$_c$  versus the classical or anomalous Dreicer field in the left (right) panel. The classical Dreicer field is plotted in black and the anomalous field is plotted in blue for E$_{c\,\, max}$ and red for E$_{c\, \, min}$. The solid line is where $\mathscr{E}_{RC}=\mathscr{E}_{D}$ and the error bars are 67$\%$ confidence intervals.\\

In the upper limit of E$_c$,  $\mathscr{E}_{RC} \ll \mathscr{E}_{D, an}$, which indicates that the return current is carried by the bulk thermal electrons. This is consistent with the drift velocity values which are lower than the threshold for the generation of current-driven instabilities \textit{but inconsistent} with the anomalous resistivity values. What if Ohm's law is invalid, i.e., the return current is not carried by the bulk thermal electrons? Then the resistivity values do not need to be as high and the Spitzer resistivity, which is the lower limit to the true resistivity in the corona, is used.\\

Figure~\ref{fig:dreicer}(a) shows that $\mathscr{E}_{RC} \geq \mathscr{E}_{D, \, \,classical}$ in 12 out of 528 spectra, which means that the entirety of the background thermal population of electrons is in the runaway regime. When $\mathscr{E}_{RC} \lesssim \mathscr{E}_{D, \, \,classical}$, only electrons with a speed greater than a critical velocity will run away. This velocity is given by $v_{cr}= \sqrt{{\mathscr{E}_D \over \mathscr{E}_{RC}}} v_{Te} $, e.g., \cite{1994PhRvL..72..645G}.\\

In the upper limit of E$_c$,  $\mathscr{E}_{RC} \ll \mathscr{E}_{D, an}$, which indicates that the return current is carried by the bulk thermal electrons. This is consistent with the drift velocity values which are lower than the threshold for the generation of current-driven instabilities \textit{but inconsistent} with the anomalous resistivity values. What if Ohm's law is invalid, i.e., the return current is not carried by the bulk thermal electrons? Then the resistivity values do not need to be as high and the Spitzer resistivity, which is the lower limit to the true resistivity in the corona, is used.\\

Figure~\ref{fig:dreicer}(a) shows that $\mathscr{E}_{RC} \geq \mathscr{E}_{D, \, \,classical}$ in 12 out of 528 spectra, which means that the entirety of the background thermal population of electrons is in the runaway regime. When $\mathscr{E}_{RC} \lesssim \mathscr{E}_{D, \, \,classical}$, only electrons with a speed greater than a critical velocity will run away. This velocity is given by $v_{cr}= \sqrt{{\mathscr{E}_D \over \mathscr{E}_{RC}}} v_{Te} $, e.g., \cite{1994PhRvL..72..645G}.\\

 To determine whether there are enough runaway electrons from the thermal plasma to carry the return current, the runaway flux density needs to satisfy the condition:
 
   \begin{equation}{j_{runaway} \over e}\, = \,n_{e} \, \gamma_{run} \, \,x_{TT} \gtrsim n_{b} \, v_{b}\label{eq:runaway} \end{equation} 
   
   where $n_b$ is the beam density in electrons cm$^{-3}$ given by equation~\ref{eq:beam_density}, $v_b={{\int f(v)\, \, v \, \,dv} \over {\int f(v)\, \, dv}}={ {\delta +1/2} \over \delta}\, \, \sqrt{2 E_c \over m}$, where $f(v)={m^2 \over 2}\,{ F(E) \over E}$ the average beam velocity in cm s$^{-1}$, n$_{e}$ is the background thermal electron density as calculated in section~\ref{sec:timeevo}; 
   
       \begin{equation}\gamma_{run}=0.35\, \nu_e\,{({\mathscr{E}_D \over \mathscr{E}_{RC}})}^{3/8}\,exp(-\sqrt{2\,{\mathscr{E}_D \over \mathscr{E}_{RC}}}\, -\, 0.25\, { \mathscr{E}_D \over \mathscr{E}_{RC}})\end{equation}
   
     is the runaway rate in s$^{-1}$ as calculated by \cite{1964PhFl....7..407K} which is a function of the ratio of the Dreicer field strength to the return-current field strength, and x$_{TT}$ is the half-length of the loop in cm. Equation~(\ref{eq:runaway}) is only valid when $\mathscr{E}_{RC} \ll \mathscr{E}_{D}$. For our sample, 370 out of 528 spectra satisfy this criterion in the upper limit of the low-energy cutoff. The various cases are summarized in table~\ref{tab:cases}.\\

\begin{figure}[bh!]
\centering
\includegraphics[width=7cm]{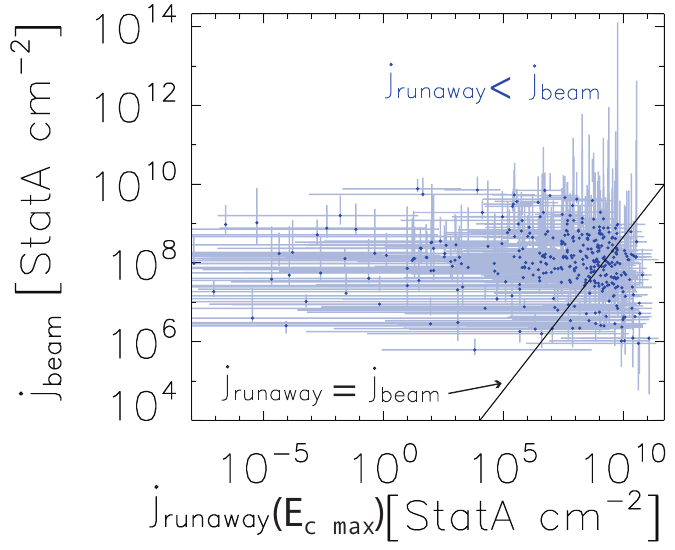}
\caption{Scatter plot of the current density of the thermal runaway electrons versus the beam current density in the upper limit of the low-energy cutoff. The only cases plotted correspond to $\mathscr{E}_{RC}\,< 0.1 \,\mathscr{E}_{D} $, for which the runaway rate could be accurately calculated. The error bars correspond to 67$\%$ confidence intervals. The solid line is where the beam and the runaway current densities are equal. There are 370 cases plotted, of which 176 (276) correspond to j$_{runaway}\geq$ j$_{beam}$ (within 67$\%$ confidence interval) and 194 (94) correspond to j$_{runaway} <$ j$_{beam}$ (within the 67$\%$ confidence interval). }
\label{fig:runaway}
\end{figure}

    Figure~\ref{fig:runaway} shows the beam current density versus the computed runaway current density for the 370 spectra with $\mathscr{E}_{RC}(E_{c\, max}) < \, 0.1 \, \mathscr{E}_{D}$.  Indeed there are many cases (176 out of 370) where the thermal runaway electrons carry the return current because $j_{runaway} \geq j_b$. The results are summarized in table~\ref{tab:cases} for both limits of the low-energy cutoff.\\
    
\begin{table}
\centering
\begin{tabular}{ccccccc}
\toprule
&$\mathscr{E}_{RC}\geq \mathscr{E}_{D}$ &  & $0.1\,\mathscr{E}_{D}<\mathscr{E}_{RC}< \mathscr{E}_{D}$  & & \multicolumn{2}{c} {$\mathscr{E}_{RC}\leq 0.1\, \mathscr{E}_{D}$} \\
\cmidrule{6-7}
&{} & {}& & & {j$_{run}\geq$ j$_b$} & j$_{run} <$ j$_{b}$\\
\midrule
E$_{c\, max}$&12 && 146 && 176 (276) & 194 (94) \\
E$_{c\, min}$& 390 & &100 && 26 & 12  \\
\hline
\bottomrule
\end{tabular}
\caption{Number of cases under each condition in the upper and lower limits of the low-energy cutoff. The numbers in parentheses represent the number of cases within 67$\%$ confidence. There are 100 more cases in the runaway regime within 67$\%$ confidence.}
\label{tab:cases}
\end{table}

In total, there are 334 time intervals out of 528 (63$\%$) for which the runaway electrons in the thermal tail could be responsible for carrying the return current if the low-energy cutoff is E$_{c\,max}$. Of those 334 spectra, 12 cases correspond to where the entirety of the thermal distribution is in the runaway regime, 146 cases where $0.1\, \mathscr{E}_{D}<\mathscr{E}_{RC}(E_{c\, max}) < \, \mathscr{E}_{D}$ (the runaway current density is significant but the runaway rate cannot be accurately calculated), and 176 where the runaway current is higher than the beam current and $\mathscr{E}_{RC}(E_{c\, max}) < \, 0.1 \, \mathscr{E}_{D}$. \\

If the low-energy cutoff is lower than E${c\, max}$, there will be more cases where the runaway current is enough to carry the return current, as shown in figure~\ref{fig:dreicer}(b) for $E_c= E_{c \,\, min}$. The majority of the cases show that the entirety of the thermal population is in the runaway regime. \\

An alternative explanation is that there is a source of turbulent (anomalous) resistivity other than the return current. For example the beam itself could become unstable and generate turbulence. This increases the effective resistivity \citep[e.g.][]{2009A&A...506.1437K}.  Another is that the return current is carried by the nonthermal beam electrons themselves. This is not possible in a 1D model, as the beam electrons are only able to move downward. However, in a 1.5 D model, the return current electric field would decelerate the parallel component of a beam electron's velocity, increasing the pitch angle of the electrons and eventually accelerating electrons with high enough pitch-angles in the opposite direction back to the looptop. Ohm's law is not valid in this case because the velocity of the upward streaming electrons is suprathermal. Similarly to the thermal runaway current, the resistivity loses its meaning because the electrons have a high enough energy to escape significant collisions. The 1.5 D model might be consistent with the results of \cite{2006ApJ...653L.149K} and \cite{2013SoPh..284..405D} that suggest a near-isotropic electron distribution. This needs to be tested for plausible beam pitch-angle distributions.\\

 In the lower limit of E$_c$,  $\mathscr{E}_{RC} \lesssim \mathscr{E}_{D, an}$, which means thermal electrons carry the return current but there can be a non-negligible population of runaway electrons, when  $\mathscr{E}_{RC}\sim\mathscr{E}_{D}$. Figure~\ref{fig:dreicer} (b) shows that $\mathscr{E}_{RC} \gtrsim \mathscr{E}_{D,\, \, classical}$  for 390 cases, which means that the entirety of the thermal background electrons is in the runaway regime and Ohm's law does not apply. The results are summarized in table~\ref{tab:cases}. \\

In conclusion, taking into account both limits of the low-energy cutoff, the return current is most likely carried by the thermal runaway electrons in most cases. Alternatively, the return current might be carried by nonthermal electrons that are pitch-angle scattered back toward the acceleration region. This explanation needs to be tested with a model that takes into account pitch-angle scattering and the return current simultaneously and self-consistently. For a few cases, as many as 56 spectra out of 528 for which the resistivity can be considered classical in the upper limit of the low-energy cutoff, and 88 spectra out of 528 in the lower limit of the low-energy cutoff, the derived resistivity values can be considered classical because the beam could be streaming down along a cool loop with temperatures as low as 2 MK. The return current in these cases is carried by the bulk thermal electrons and Ohm's law is valid. Finally, there are at most 79 cases in the lower limit of E$_{c0}$, where the return current could be unstable to either the EIC, IA or Buneman instability, depending on the ratio of the electron to ion temperature. \\

\section{Return-current heating and constraints on the low-energy cutoff}
\label{sec:heating}
\subsection{Are all flares with a strong spectral flattening consistent with return current losses?}
\label{energy_total}
We have found that the RCCTTM provides acceptable fits to RHESSI spectra with strong breaks and that their time evolution is sometimes smoother than obtained for the standard collisional thick-target model fits (section ~\ref{rccttm_cttm} and figure~\ref{fig:chisq}). We have also found that the return current can be carried by the thermal runaway electrons, which do not contribute to the Joule heating because these electrons do not experience collisions. In this section we answer the following questions: Is the heating from return-current losses due to the potential drop consistent with observations of thermal emission? Using the heating signatures, is it possible to differentiate between a flattening due to the return-current potential drop and the low-energy cutoff?\\

The spectra from both the 2005-Jan-19 and -20 flares were well fitted with the RCCTTM. We again focus on the time intervals with a significant potential drop: the last two HXR peaks during 2005-Jan-19 from 08:24 UT to 08:29 UT, and the first HXR peak during the 2005-Jan-20 event between 06:42:42 UT and 06:49:00 UT. Unfortunately, as both the return current model and a single (as for 2005-Jan-19) or double (as for 2005-Jan-20) power-law fits provide comparable fits, we cannot conclude from the fits alone whether return-current losses are significant. However, the time evolution of the fit parameters shows that the return current model works better for 2005-Jan-20 (see section~\ref{rccttm_cttm}). The two electron distributions, with or without a potential drop, will have different heating signatures: Return-current losses will primarily heat the corona as the electric field decelerates the nonthermal electron beam, and an electron distribution with a low-energy cutoff as high as $\sim$120 keV, as in 2005-Jan-19, reaches the footpoints without any significant energy losses. An electron of energy 120 keV only loses 2 keV through Coulomb collisions in a 5 Mm loop of density 10$^{11}$ cm$^{-3}$. In addition, the energy flux density is insufficient to heat the corona through chromospheric evaporation, as discussed in \cite{2009ApJ...699..917W}, who assume collisional losses.\\

The total nonthermal energy lost between the looptop and the thick target is calculated through: \begin{equation}E_{nth}(t) =\, \int_{0}^{t} \Delta\mathscr{F}\, \, dt \label{eq:nth} \end{equation} where t is the time since the beginning of the HXR emission, and the energy flux difference $\mathscr{F}$ between the looptop and the thick target is calculated as follows:  \begin{equation} \Delta\mathscr{F}= \mathscr{F}_{0}- \mathscr{F}_{TT}= \int_{E_c}^{\infty} F(E_0) E_0 dE_0 - \int_{E_{c \, TT}}^{\infty} F(E, x_{TT}) E \, \, dE \label{eq:energy_loss} \end{equation}

where $\mathscr{F}_0$ and  $\mathscr{F}_{TT}$ are the injected energy fluxes at the looptop and the thick target in erg s$^{-1}$, respectively.\\
$F(E_0)=(\delta -1)\, F_0\, E_c^{\delta-1} \,E_0^{-\delta}$ and $F(E,x_{TT})= (\delta -1)\, F_0\, E_c^{\delta-1} \,(E+e\,V_{TT})^{-\delta}$ are the electron distributions at the looptop and thick target in electrons s$^{-1}$ keV$^{-1}$, respectively. $F_0$ is the total injected electron number flux.\\

After calculating the integrals in equation~\ref{eq:energy_loss}, equation~\ref{eq:nth} becomes, for E$_{c\, max}$, \begin{equation}E_{nth}= \int_0^t e\,V_{TT} \, F_{0}\, dt \label{eq:enth}\end{equation}

\begin{figure}
\centering
\includegraphics[width=18.5cm]{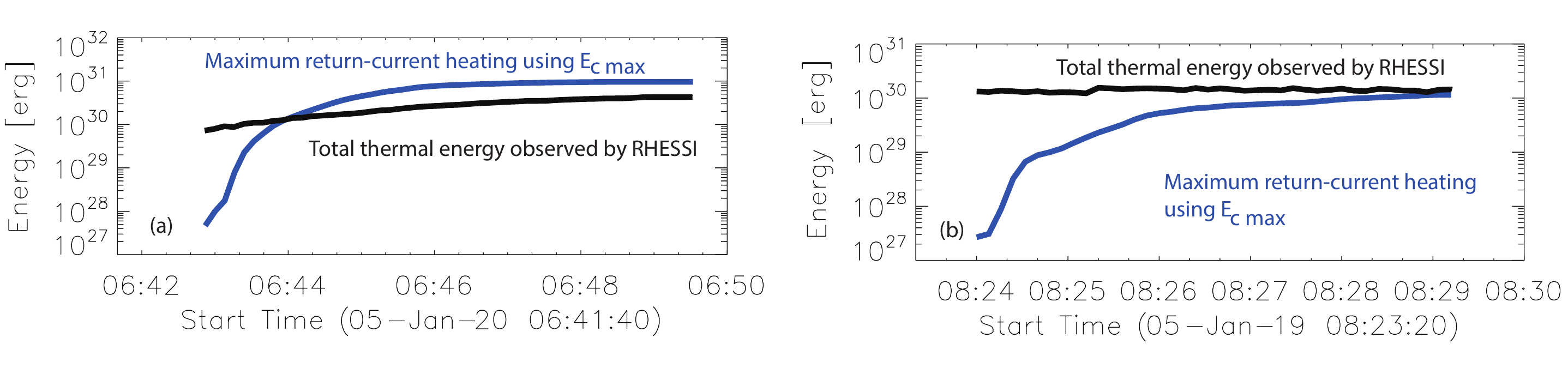}
\caption{Time evolution of the total thermal energy deduced from RHESSI observations in black, and the maximum nonthermal energy lost between the looptop (LT) and thick target (TT) due to the potential drop using the upper limit of the low-energy cutoff in blue. The left and right panels correspond to the first HXR peak in 2005-Jan-20 and the last HXR peak in 2005-Jan-19, respectively.   }
\label{fig:energy}
\end{figure}

Figure~\ref{fig:energy} shows the time evolution of the maximum derived nonthermal energy deposited in the corona due to return current losses in the upper limit of the low-energy cutoff, assuming all of this energy goes into heating the coronal plasma, as calculated by equation~\ref{eq:enth}. These curves are shown in blue for both the 2005-Jan-19 and -20 flares. The total thermal energy in the corona is plotted in black, and is given by $2\, n_{e}\, k T\, \Omega$, where $\Omega$ is the thermal volume in the corona as calculated in section ~\ref{source sizes}, T the best-fit temperature, and $n_{e}$ the thermal background density. \\

 Figure~\ref{fig:energy}(a) shows that the thermal emission rises by a factor 8 from 06:43 UT to 06:50 UT.  The lost nonthermal energy is about 2 orders of magnitude lower than the thermal energy at 06:43 UT but becomes equal to the thermal energy 06:44 UT, yet the thermal energy is rising. This indicates that something other than the return current potential drop is heating the corona during the first minute. This could be chromospheric evaporation, or the low-energy cutoff is lower than the upper limit E$_{c\, max}$. If the low-energy cutoff is lower than E$_{c \, max}$, then the electron flux becomes higher, which results in more heating. Additionally, collisional losses become more important the lower the low-energy cutoff, and so does the thermalization of electrons, which heats the plasma and decreases the electron flux from the looptop to the thick target. After 06:44 UT, the nonthermal energy in the upper limit of E$_c$ seems sufficient to heat the coronal plasma. The spectral flattening during 2005-Jan-20 event is consistent with return current losses. Note that the contribution to the heating of the return current could be higher if the resistivity is higher than the Spitzer values derived using the fitted temperature. For example, if the beam is streaming in a cooler loop than the fitted RHESSI temperature, i.e., if the observed RHESSI thermal emission is from previously heated plasma in different loops.  Another way the heating could be higher is if the resistivity is anomalous due to turbulence other than from a current-driven instability.\\

  Since the return current could be carried by runaway electrons from the tail of the thermal distribution, as demonstrated in section~\ref{dreicer section}, these electrons do not contribute to the Joule heating. However there is a minimum heating produced due to collisional friction (Spitzer resistivity) when the thermal background plasma is not entirely in the runaway regime. A rigorous calculation of the heating needs to be performed in the full range of possible low-energy cutoffs. This heating will also depend on the position along the loop because the electron flux may decrease due to thermalization of lower energy electrons, and the runaway current depends on the position along the loop as inferred from equation~(\ref{eq:runaway}). Collisional losses when the low-energy cutoff is low enough need to be taken into account. It is possible that return currents do not contribute significantly to the heating of the corona if the thermal runaway electrons are dominant over bulk thermal electrons. \\

Figure~\ref{fig:energy}(b) shows that the thermal energy is constant for the 5 min duration of the HXR emission after 08:24 UT, and that the nonthermal energy loss due to the potential drop is on the same order of magnitude as the total thermal energy. The non-thermal energy has been calculated for the last HXR peak starting at 08:24 UT. This could indicate that the return current is mostly carried by thermal runaway electrons which would not contribute to the heating, keeping the thermal emission constant. \\

A closer look at the beam and plasma parameters from the \href{https://hesperia.gsfc.nasa.gov/collaborate/malaouia/public_html/movie4html/uncert20050119.pdf}{2005-Jan-19} show that the return current is stable to the generation of EIC and IA instabilities. The return current electric field is an order of magnitude lower than the classical Dreicer field, which allows us to calculate the runaway rate. The runaway current is more than an order of magnitude higher than the beam current, which means that runaway electrons in the thermal tail carry the return current, resulting in insignificant heating from the potential drop. \\
 
 It is not possible to conclude whether the absence of additional heating in 2005-Jan-19 after 08:24 UT is due to a high value of the low-energy cutoff ($>$100 keV) or the possibility that the return current is carried by runaway electrons in the thermal tail. It is important to note, from a theoretical point of view, that return currents should be present in all cases as they are essential to keep the beam from being pinched off.

\subsection{Can we better constrain the low-energy cutoff?}
\label{elow}
Starting with the assumption that the strong flattening in an X-ray spectrum is completely due to flattening in the electron distribution due to return current losses, we have obtained an upper limit for the low-energy cutoff consistent with potential drop flattening and a lower limit given by the analytical result derived by \cite{2015ApJ...809...35K}. However, we can constrain the lower limit of the low-energy cutoff to a higher value than $\delta$ k T, in some cases, considering a physical inconsistency. Two physical inconsistencies provided constraints on the lower limit of the low-energy cutoff, namely (1) a derived resistivity lower than the Spitzer resisitivity, and (2) a beam density higher than the background density.
\\

It is unphysical for the resistivity in the corona to be lower than the Spitzer resistivity of the highest temperature component. Evidently, there could be a hotter component in the corona with lower emission measure that was not observed with RHESSI, because the dynamic range of RHESSI only allowed the highest intensity source to be observed. We tested whether adding another thermal component to the fit model improves the fit and whether the potential drop is affected. None of the resistivities could be explained by a second, hotter thermal component in the spectra. In addition, adding another thermal component kept the potential drop unaffected in all events, and only improved the fits in the \href{https://hesperia.gsfc.nasa.gov/collaborate/malaouia/public_html/movies4html/20020226/movie.html}{2002-Feb-26} event.\\

\begin{figure}
\centering
\includegraphics[width=8.5cm]{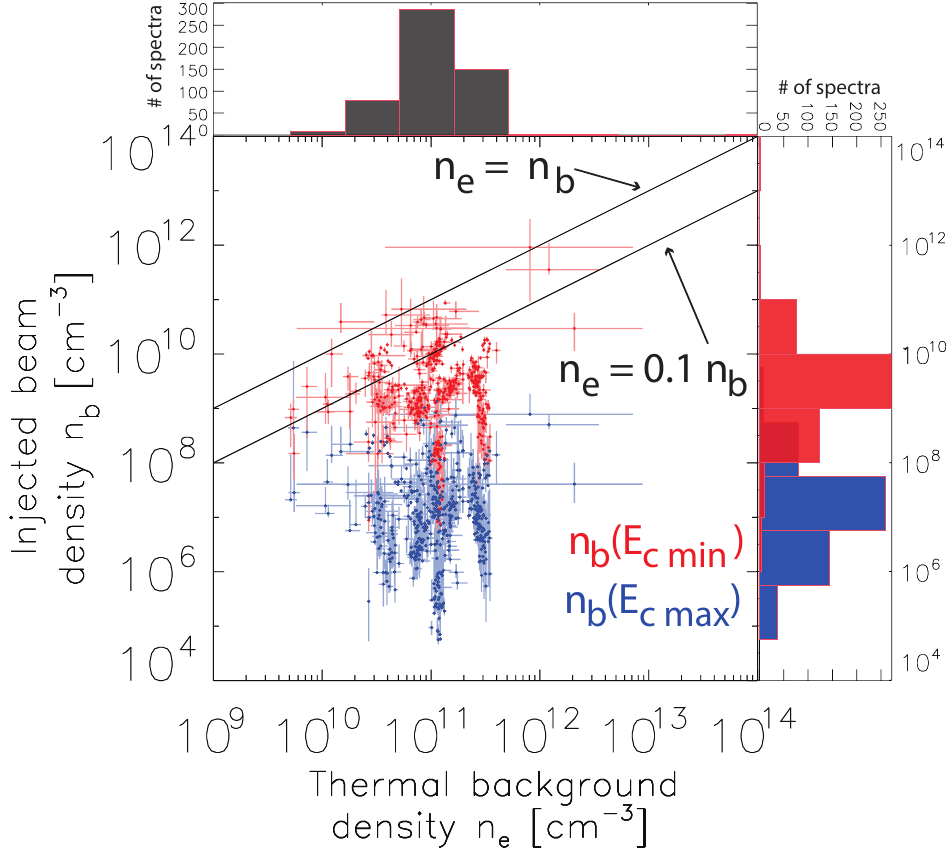}
\caption{ Beam density in the upper (blue) and lower limits of the low-energy cutoff versus the thermal background density. The error bars are 67$\%$ confidence intervals.}
\label{fig:density}
\end{figure}

There is no observational evidence for coronal temperatures higher than $\sim$50 MK \citep[][]{2010ApJ...725L.161C}. Calculating the temperature assuming the derived resistivity from Ohm's law in these apparently sub-Spitzer cases to be classical Spitzer, we obtain temperatures between 42 and 65 MK if we consider only spectra with a potential drop higher than 10 kV. There are clearly cases with too high a temperature to be considered coronal. These correspond to spectra that are well fitted with a single power-law and sharp low-energy cutoff lower than or equal to the transition energy between the thermal and nonthermal components. Hence, the lower limit of the low-enegy cutoff should be higher than $\delta$ k T.\\

Another requirement of the model is that the background density be higher or much higher than the density of the beam because there needs to be a balance between the return and direct currents such as: n$_b v_b = n_e v_d$. Since $v_b \gg v_d$, the background density needs to be higher than the beam density. Figure~\ref{fig:density} shows that the beam density in the upper limit of the low-energy cutoff is always lower than the background density by at least an order of magnitude. In the lower limit of the low-energy cutoff, however, the beam density is higher than the background's in 4 cases and $n_b > 0.1\, n_e$ in 86 out of 528 cases. This indicates that the lower limit of the low-energy cutoff is too low. \\

These results were derived assuming a filling factor of 1. Equation~(\ref{eq:density_beam}), shows that the electron number density is proportional to the electron flux density $F_{e0}\,=\, F_0 \over {A_{FP}\,\, f_1} $, where $F_0$ is the electron flux in electrons s$^{-1}$, A$_{FP}$ is the cross-section of the flare loop as calculated in section~\ref{source sizes} and $f_1$ is the filling factor of the area. Equation~\ref{eq:density_bkg}, gives the thermal background density such as $n_{e} = {\sqrt{{EM} \over {\Omega \,\, f_2}}}$, where $\Omega$ is the thermal volume as calculated in section~\ref{source sizes} and $f_2$ is the filling factor of the thermal volume. The two filling factors $f_1$ and $f_2$ are equal because in a 1D model, the volume can be thought of as a filamented cylinder and the same filaments go through the cross-section of the loop. Therefore ${n_b \over n_e} \propto f^{-1/2}$.\\

 The filling factor has been deduced to be as low as $10^{-3.7}$ from cooling times of $\sim$ 4500 RHESSI microflares \cite{2011ApJ...736...75B}. If we take $f=10^{-4}$, the ratio $n_b \over n_e$ increases by two orders of magnitude. The result is that in the lower limit of the low-energy cutoff there are 503 out of 528 cases where $n_b >\, 0.1\,\, n_e$, and in the upper limit of the low-energy cutoff, there are 57 out of 528 cases where $n_b >\, 0.1\,\, n_e$. If the filling factor is $f=10^{-2}$ this latter drops to 10 out of 528 cases. In summary, if the filling factor is lower than 1, the lower limit of the low-energy cutoff becomes too low for most of the cases. The lower the filling factor the more cases violate the $n_b< 0.1\, n_e$ condition, even in the upper limit of the low-energy cutoff.

\section{Summary}
\label{summary}
Main result: The co-spatial return-current collisional thick-target model (RCCTTM) provides acceptable fits to HXR flare spectra. 1154 8-s spectra from 19 flares were studied. Acceptable fits are defined by (1) a reduced $\chi^2$ with values between 0.49 and 1.48 (which corresponds to the 95$\%$ interval of reduced $\chi^2$ distribution with 40 degrees of freedom) during the HXR peak times; (2) a smooth time evolution of the fit parameters. When the potential drop is negligible the RCCTTM is equivalent to the standard collisional thick-target model, where the injected spectrum in the corona reaches the thick target unchanged. Those time intervals with a potential drop less than 10 kV were not included in the statistical analysis, leaving 528 spectra.\\

In general, whenever the standard thick-target model with a sharp low-energy cutoff and a single or double power-law provides spectral fits within the 95$\%$ confidence level to HXR spectra, the return current model also provides an acceptable fit. The only exception we found was 2002-Aug-22 (see section~\ref{stat}). \\

The return current model provides a physical explanation for the strong breaks in our sample. These breaks were chosen because the difference in spectral indices is higher than what other mechanisms such as non-uniform ionization and isotropic Compton back-scattering can explain. Also, the return current and the co-spatial electric field that drives it must be present, because the background plasma is conductive and acts to locally neutralize the beam, preventing it from being pinched off, as described in section~\ref{introduction}. \\

Two different Soft-Hard-Soft (SHS) behaviors are observed. One attributed to the acceleration region (anticorrelation of the electron spectral index and the HXR flux), and the other to the return current potential drop (correlation of the potential drop and HXR flux). The 2005-Jan-19 flare is associated with SHS behavior related to the acceleration mechanism on each separate HXR peak before 08:24 UT, and a potential drop related  SHS behavior after 08:24 UT. During the first HXR peak of the 2005-Jan-20 flare, the SHS behavior corresponds to the return-current potential drop.\\

Under the assumption that the flattening in HXR spectra is entirely due to return-current losses, we have deduced the conclusions summarized below. 
The following conclusions are obtained using E$_{c\,max}$, the upper limit of the low-energy cutoff derived from the RCCTTM fit.

\setlength{\parindent}{0.5cm}
\begin{enumerate}
 \setcounter{enumi}{0}
\item{The resistivity in the corona must be enhanced if the return current is enterely carried by drifting thermal electrons. The derived resistivity is equal to or higher by up to five orders of magnitude than the classical Spitzer resistivity at the fitted temperature. The mean value is two orders of magnitude higher than Spitzer resistivity.}

\item{The return current is likely stable to current-driven instabilities: The drift velocity is lower than the threshold for the well-studied electrostatic ion cyclotron and ion acoustic current-driven instabilities to arise. Thus, the enhanced resistivity is unlikely to be due to the return current itself. The simple 1D model where the return current is stable and carried by the bulk background electrons does not explain the anomalous resistivity values derived.} 

\item{The computed current of runaway electrons accelerated out of the thermal plasma by the return-current electrons is higher than or equal to the electron beam current in 334 cases out of 528 (63 $\%$). Note that if the 67$\%$ uncertainty is taken into account, 434 (82$\%$), of the cases could have a runaway current that balances or formally exceeds the beam current. Hence, runaway electron carrying the return current is the preferred explanation for most of the spectra. The resistivity does not need to be enhanced and the return current is stable to the current-driven instabilities.}

\item{The beam of electrons could be streaming in a loop cooler than the temperature derived from the  RHESSI fits, since the hot plasma seen with RHESSI may have been previously heated in different loops. However, this can only explain 56 out 528 cases, where the resistivity is less than that of a 2 MK plasma. (cf. figure~\ref{fig:resistivity} and section~\ref{resistivity section}).  }

\item{An alternative scenario is that the return current is carried by a portion of the original nonthermal electrons in the beam. This requires relaxing the 1D assumption to a 1.5 D model which takes into account a distribution of pitch-angles. This also solves the inconsistency between the deduced anomalous resistivity and stability of the return current to current-driven instabilities, and is possibly consistent with the deduction of quasi-isotropic beams by \cite{2006A&A...446.1157K}. This would reduce or eliminate the need for anomalous resistivity. Note that is it also possible that the anomalous resistivity is due to an external source of turbulence, not driven by the return current.}

\item{There are enough electrons in the background plasma to carry the return current: The density of the nonthermal beam is lower in most cases by at least an order of magnitude than the background density, assuming a filling factor of 1 for the SXR and HXR sources.}

\end{enumerate}

The following conclusions are obtained using $E_{c\,min}=\, \delta\, kT$, the lower limit of the low-energy cutoff derived from the warm thick-target model.

\begin{enumerate}
 \setcounter{enumi}{6}

\item{All or most of the thermal population of electrons is in the runaway regime. The return current electric field is higher than the classical Dreicer field in 390 (74$\%$) cases. If the 67$\%$ uncertainty is taken into account, 480 (91$\%$) of the cases could have the entirety of the thermal distribution in the runaway regime (Table~\ref{tab:cases}). All the cases where the runaway current could balance the beam current, in the lower limit of E$_c$, is 514 (97$\%$).}

\item{The return current is likely stable to current-driven instabilities and the return current is carried by runaway electrons from the thermal tail. All of the cases where the drift velocity is higher than the threshold for EIC and IA instabilities are associated with low-energy cutoff values that are too low. These low-energy cutoff values produce an energy in the corona from Joule heating at least 3 orders of magnitude higher than the observed RHESSI thermal energy.}

\item{The range of low-energy cutoffs can be better constrained. (1) The derived resistivity is lower than the classical Spitzer resitivity in 9 cases. This is unphysical and alternatively might indicate the presence of a higher temperature component with temperatures as high as 65 MK. Note that adding a second thermal component did not improve the fits except for the 2002-Feb-26 flare and did not affect the values of the potential drop. (2)The density of the nonthermal beam is higher than a tenth of the thermal background plasma in 90 cases assuming a filling factor of 1 for the SXR and HXR sources.}

\end{enumerate}

Main conclusion: In 1D the return current is not likely to be primarily carried by the bulk thermal electrons. Rather, runaway electrons from the tail of the thermal distribution are most likely responsible for carrying the return current. Future work will include (1) the calculation of local heating rates as a function of the low-energy cutoff and the fraction of the thermal electrons in the runaway regime, (2) calculation of the fraction of the return current consisting of beam electrons that have been pitch-angle scattered through 90$^{\circ}$ so that they propagate back toward the acceleration region.\\


\end{document}